\documentclass[11pt]{article}

\usepackage{amsmath}
\usepackage{geometry} % See geometry.pdf to learn the layout options. There are lots.
\usepackage{graphicx} 
\usepackage{color} 
\usepackage{amssymb, amsthm} 
\usepackage[]{natbib}
\usepackage{hyperref} 

\usepackage{booktabs}
\usepackage[table,xcdraw,dvipsnames]{xcolor}
\usepackage{soul}
\usepackage{caption}
\usepackage{subcaption}
\usepackage{lscape}
\usepackage{setspace}
\usepackage{setspace}
\usepackage{float}
\usepackage{comment}

\numberwithin{equation}{section}
\def\ind{\ensuremath{\stackrel{\text{ind}}{\sim}}}%
\def\Pr{{\ensuremath{\mathbb P}}}%
\def\Exp{{\ensuremath{\mathbb E}}}%
\let\hat\widehat%
\let\tilde\widetilde%
\def\given{{\,|\,}}
\def\argmax{\mathop{\rm arg\,max}\limits}%
\def\argmin{\mathop{\rm arg\,min}\limits}%

\title{Modeling Urban Crime Occurrences via Network Regularized Regression}

\author{Elizabeth Upton$^1$ and Luis Carvalho$^2$\\[1em]
\normalsize 1. Williams College, Department of Mathematics and Statistics\\
\normalsize 2. Boston University, Department of Mathematics and Statistics}
\date{}

\date{}

\begin{document}
\maketitle

\begin{abstract}
Analyses of occurrences of residential burglary in urban areas have shown that
crime rates are not spatially homogeneous: rates vary across the network of
city streets, resulting in some areas being far more susceptible to crime than
others.  The explanation for why a certain segment of the city experiences high
crime may be different than why a neighboring area experiences high crime.
Motivated by the importance of understanding spatial patterns such as these, we
consider a statistical model of burglary defined on the street network of
Boston, Massachusetts.  Leveraging ideas from functional data analysis, our
proposed solution consists of a generalized linear model with vertex-indexed
covariates, allowing for an interpretation of the covariate effects at the
street level.  We employ a regularization procedure cast as a prior
distribution on the regression coefficients under a Bayesian setup, so that the
predicted responses vary smoothly according to the connectivity of the city.
We introduce a novel variable selection procedure, examine computationally
efficient methods for sampling from the posterior distribution of the model
parameters, and demonstrate the flexibility of our proposed modeling structure.
The resulting model and interpretations provide insight into the spatial
network patterns and dynamics of residential burglary in Boston.

\noindent 
\\[1em]
\textbf{Keywords}: graph Laplacian, network inference, residential burglary

\end{abstract}

\section{Introduction}
\label{s:intro}

\begin{figure}
\centerline{\includegraphics[scale = 0.65]{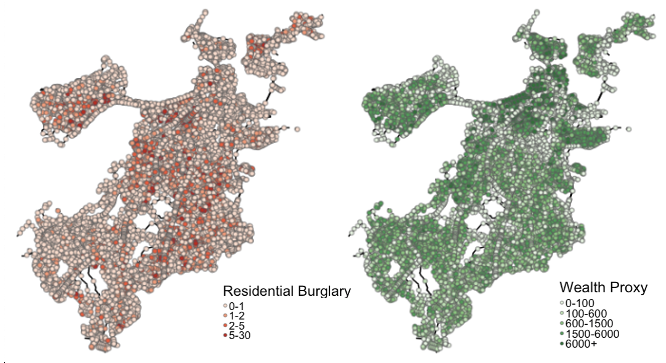}}
\caption{Residential burglary occurrence counts (left) and average wealth
estimate in thousands of dollars (right) for street segments in the network of Boston, MA.}
\label{fig:bostondescriptive}
\end{figure}

While the modeling and forecasting of crime have been of interest to law
enforcement and governmental agencies for some time, the recent availability of
large crime datasets in urban areas has substantially increased and advanced
these efforts.  Here, we are interested in a specific type of crime,
residential burglary, legally defined as ``the act of breaking and entering a
building with the intent to commit a felony"~\citep{garner2001}.  Using the GPS
coordinates of reported burglaries~\citep{opendata}, we mapped each crime
occurrence (from June 2015 through December 2019) to its street segment.  We
then defined the dual network graph of Boston~\citep{dual},  where street
segments are represented as nodes and edges are placed between any two which
share an intersection.  Figure~\ref{fig:bostondescriptive} pictures the
network, with each vertex color-coded to indicate our attribute of interest,
counts of residential burglary.

\begin{figure}
\centering
\includegraphics[scale=.25]{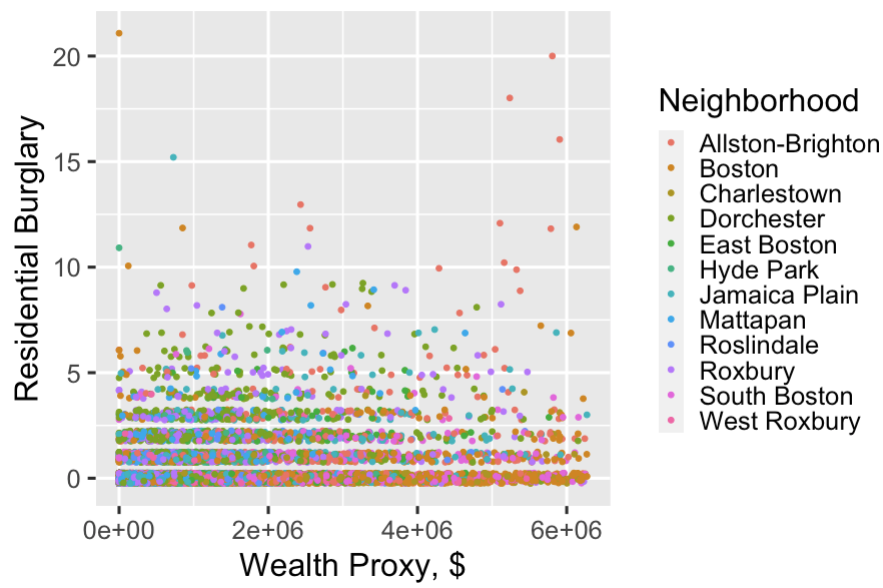}
\caption{Counts of residential burglary (slightly jittered) versus wealth estimate by neighborhood for each street segment; illustrates the nonconstant effects of wealth on crime.  To enhance the visualization, we exclude street segments with wealth estimates greater than the 95th percentile (\$6,272,010). }
\label{fig:corr}
\end{figure}

A naive approach to modeling crime counts across the network would entail
identifying a set of predictor attributes describing each street segment and
performing count regression.  For example, if $Y_v$ and $\mathbf{x}_v$ are the
crime occurrence counts and covariate attributes at vertex $v$, we could
suppose
$Y_v \ind \text{\sf Po}[\exp(\textbf{x}_v^T \beta)]$.
However, this specification assumes that the effects $\beta$ are constants
across the network and empirical evidence suggests this is not the case.  For
example, Figure~\ref{fig:bostondescriptive} displays gross tax
information for the city of Boston in 2015.  Comparing this attribute to the
counts of residential burglary, we see that in some areas of
the city lower taxes, identifying poorer communities, correspond to larger
crime rates; however, in other locations, higher taxes indicative of wealth,
correlate with higher crime rates.  This is illustrated more clearly in
Figure~\ref{fig:corr} where we see counts of residential burglary versus the
wealth proxy by street segment; the correlation between these two variables is
0.15. Thus, even when including informative covariates, a regression with
constant effects cannot adequately explain the variability in burglary
occurrences across a city.

Additionally, the existence of crime ``hot spots'', areas of concentrated crime
counts, is widely acknowledged in crime theory literature~\citep[see,
e.g.,][]{eck2005}. These zones can be identified visually in
Figure~\ref{fig:bostondescriptive}; for example, in the southern region of the
northwest peninsula neighborhoods, Allston and Brighton. Many current crime
mapping algorithms focus on identifying these hot spots and estimate future
crime risk based on past crimes~\citep{bowers,kim2018}.  However,
interpretative models providing probabilities of hot spot formulation
accounting for and explaining non-homogeneous effects are lacking.

A street network approach to crime modeling is beneficial, as it allows for the
creation of a high-resolution spatial model.  The measurements of distance
utilized by the model are based on street length, connectivity, and
walkability, naturally incorporating environmental and structural barriers
within the urban landscape.  Furthermore,  the connection between urban
configurations and crime patterns aligns with key criminological theories.  For
example, crime pattern theory posits that criminals often target properties
they encounter during non-criminal activities, linking the risk of a certain
location being victimized to travel patterns that arise from every day
activity~\citep{davies}.  The hierarchical structure of city streets, with
highly connected segments experiencing increased pedestrian and vehicle
traffic, is captured in the network representation of the city and can be
leveraged by a regression model that incorporates network information.  Current
approaches for modeling the spatial distribution of crime, such as point
processes or areal/aggregate methods~\citep{mohler, balocchi} utilize Euclidean
distance metrics, do not capture patterns at the street segment level, or focus
primarily on forecasting as opposed to interpreting covariate effects.

Our proposed method aims to capture both gradual variations in crime explained
by predictor attributes and abrupt variations attributed to hotspots.  Most
significantly, the model does not only identify areas of the city with extreme
levels of residential burglary, but it identifies how factors contributing to
these crimes change across the city. With this information, local law
enforcement can direct crime prevention efforts to narrowly defined regions of
the city and identify area specific interventions aimed at decreasing the
occurrence rate of residential burglary.

\subsection{Outline and Main Contributions}
The outline of this paper is as follows: next, in Section~\ref{sec:model}, we
introduce a regression model based only on the structure of a network.  We then
extend these ideas to include predictor attributes and cater our model to the
needs of crime prediction.  We also briefly discuss prior and related
methodological work. In Sections~\ref{sec:priorinference}
and~\ref{sec:modelinference}, we provide guidelines for eliciting hyper-prior
parameters and outline how to fit the model. In Section~\ref{sec:analysis}, we
illustrate and compare the performance of our model on the Boston residential
burglary data and simulated crime data. Finally, we conclude with a brief
summary and discussion of future extensions in Section~\ref{sec:conclusions}.

Our methodological contributions are outlined below, and details of each are
discussed in the forthcoming text.

\begin{itemize}
\item We construct a \textit{residential} street network of Boston by
restricting the vertex set to street segments that abut residential parcels
($n$ = 12,763).  To maintain the connectivity of the entire vertex set on the
reduced set, we utilize the Schur complement of the block of non-residential
intersections on the full network Laplacian matrix.
%This allows us to marginalize out the non-residential intersections where we would expect zero occurrences of residential burglary

\item We address the issue of effect nonhomogeneity by vertex indexing our
regression coefficients, allowing them to vary across the network.
To accomplish this, we examine the eigendecomposition of the network Laplacian
and adopt basis expansions of varying sizes composed of subsets of the
eigenvectors of the Laplacian matrix for each coefficient.

\item To avoid overfitting the model we impose smoothness on the linear
predictor using a penalty based on a discrete differential operator induced by
the network Laplacian matrix~\citep{ramsay2005}.
%In this manner
%we incorporate information from both the network topology and meaningful
%predictors into our regression model.

\item To perform variable selection and determine the rank of each basis
expansion we employ a novel sequential spike-and-slab prior on the basis
expansion coefficients~\citep{george1993}.
%That is, we adapt the common Bayesian variable selection
%technique to address the dependence structure induced by determining a rank
%rather than a binary ``include/do not include'' decision for each possible
%expansion
In the crime application, the rank of the basis expansion for each
predictor variable allows an interpretation on the magnitude and variation of
the predictor effect across the city.

\item Addressing abrupt changes in crime rates, we use a latent network-indexed
indicator to identify residential burglary hot spots. The indicator attribute
assigns to each intersection its hot spot status, and is designed to vary
smoothly over the network in the same manner as the other network predictors.
%In effect, this specification results in a zero-inflated
%count regression since the model now defines a ``background'' crime rate and a
%separate hot zone crime rate for each intersection.

\item  The proposed hierarchical model is formalized in a Bayesian setup with
Gaussian priors on the parameter sets. We use a block Gibbs
sampler~\citep{monte} to draw samples from the posterior distribution of the
model parameters.  We also propose a computationally efficient
expectation-maximization (EM) algorithm~\citep{dempster1977} to find maximum
\emph{a posteriori} estimates of the parameters.

%Due to the potentially large parameter space, where
%Gibbs sampling may be slow or even fail to converge, we propose a
%computationally efficient expectation-maximization (EM)
%algorithm~\citep{dempster1977} to find maximum \emph{a posteriori} estimates
%of the parameters and use these values as a starting point in the Gibbs
%sampler. This EM procedure is also used to elicit suitable priors based on
%prediction error minimization.
\end{itemize}

\section{Proposed Model and Related Work}
\label{sec:model}
Our main methodological tool is \emph{network regression}: given a network
where vertices are connected by weighted edges, we observe vertex indexed data
in the form of vertex attributes, and, as common in many applications, we
distinguish between a response attribute of interest and a set of predictor
attributes. To regularize the response attribute, we let the edge
weights capture a measure of similarity between predictor attributes, as
discussed shortly. Examples of representations of this data structure, beyond urban crime,
include the infection status of individuals in a network of injection drug
users given their drug use habits and other covariates such as age and gender;
the political party affiliation of web blog authors in a network of hyperlinked
connected blogs given the distribution of post topics and readership
ideological inclinations; and the functional classes of proteins in a network
of protein-protein interactions given gene pathway and other biological
information~\citep{snapnets}.
%Our main interest is then to model the response attribute using regression, but
%in a way that explores the vertex similarity information in the network.
%Furthermore, we want to address effect nonhomogeneity, that is, a situation
%where the covariate effects vary across the network.  Before discussing our
%proposed solution we describe the application motivating our efforts.

We further abstract our problem of interest as follows: consider a weighted
graph $G = (V, E, w)$ with vertex set $V$, edge set $E$, and positive weights
$w$, that is, $w_{ij} > 0$ whenever $(i, j) \in E$ for all $i, j \in V$ and
$w_{ij} = 0$ otherwise.
We wish to regress a response $Y$ on a set of predictor attributes
$X$ using a generalized linear model: for each vertex $v \in V$,
$Y_v \given \beta \ind \text{\sf F}[g^{-1}(\mathbf{x}_v^\top \beta(v))]$,
where $\sf F$ belongs to an exponential family and $g$ is a link function, and
following our discussion, the network effects $\beta$ are also vertex indexed.
% TODO: discuss in conclusion
%In general, we could regress a response attribute $Y$ on the
%network-indexed coefficients,
%$Y_v \given \beta \ind \text{\sf F}[g^{-1}(\mathbf{x}_v^\top \beta(v))]$,
%where {\sf F} belongs to the exponential family and $g$ is a link.

\subsection{Single-intercept model}
Let us assume, for now, that we have a single-intercept model.  Also, to
directly apply the methodology to the crime application, we presume a count
response:
$Y_v \given \beta \ind \text{\sf Po}[\text{exp}(\beta(v))].$
To avoid overfitting, we impose smoothness on $\beta$ through an informative
prior that measures the roughness of $\beta$ using a differential operator
$M_w$ and a roughness penalty $\lambda > 0$:
$\beta \sim N(0, \lambda^{-1} {(M_w^\top M_w)}^{-})$.
The maximum \emph{a posteriori} (MAP) estimate $\hat{\beta}$ for $\beta$ is
then
\[
\begin{split}
\hat{\beta} &= \argmax_{\beta} \Bigg\{\ell(\beta; Y) - \frac{\lambda}{2}
\beta^\top M_w^\top M_w \beta \Bigg\} \\
&= \argmin_{\beta} \Big\{\mathcal{D}(Y, g^{-1}(\beta)) +
\lambda \beta^\top M_w^\top M_w \beta \Big\},
\end{split}
\]
where $\ell$ is the model log-likelihood and $\mathcal{D}$ is the model
deviance.

We take $M_w$ to be the oriented weighted incidence matrix of the network: for
$e = (i, j) \in E$, $M_{w,ei} = \sqrt{w_{ij}}$, $M_{w,ej} = -\sqrt{w_{ij}}$ and
$M_{w,ev} = 0$ for all $v \in V$, $v \neq i$ and $v \neq j$.
Thus $L_w = M_w^\top M_w = D_w - W$ is the weighted graph Laplacian with
$W = [w_{ij}]$, the weighted adjacency matrix, and
$D_w = \text{Diag}_{i \in V} \{\sum_{j \in V} w_{ij}\}$, a
diagonal matrix with the weighted degrees. In this sense, the prior on $\beta$
is similar to an intrinsic conditional autoregressive
model (ICAR; ~\citeauthor{besag}, \citeyear{besag}), but with adjacency weights.
Note that $L_w \mathbf{1}_{|V|} = 0$ and so $L_w$ is rank deficient, requiring
the use of the generalized inverse $L_w^-$ when defining the prior.
Our choice of $M_w$ is deliberate and designed to exploit information found in
the structure of the network to inform our model.
% exploit basic identities in
%spectral graph theory to leverage the network topology and weights in the
%prior for regularization.
%This approach has been followed before in the
%context of regularized least squares and kernel regression in
%graphs~\citep{smola2003,belkin2004}, where the log prior is interpreted as a
%penalty for a (log) Gaussian likelihood.
%The edge weights
%are assumed to indicate vertex affinity and adjacent or close vertices in
%the network are similar on some characteristic level relevant to the vertex
%attribute of interest; we discuss choice of weights in
%Section~\ref{sec:priorinference}.
In fact, expanding the log prior we have
\[
\beta^\top M_w^\top M_w\beta = \beta^\top L_w \beta =
\sum_{(i, j) \in E} w_{ij}{(\beta(i) - \beta(j))}^2,
\]
that is, the term penalizes the weighted sum of squared differences for
coefficients of adjacent vertices in the network~\citep{kolaczyk2009}.
Thus, as usual in Bayesian inference, we seek estimates of $\beta(v)$ that
balance representativeness with respect to our observed data $Y$ and $X$ in
the likelihood with smoothness with respect to the network topology in the
prior.

The network effects $\beta$ can be conveniently represented using a basis
expansion with respect to the eigenvectors of the operator $L_w$, a common
approach in functional data analysis~\citep{ramsay2005}. More specifically,
since $L_w$ is symmetric it realizes an eigen-decomposition
$L_w = \Phi \Xi \Phi^\top$, and we can take $\tau$ eigenvectors to represent
$\beta$ as $\beta = \Phi_{1:\tau} \theta$, where $\Phi_{1:\tau}$ contains the
first $\tau$ eigenvectors of $L_w$, ordered by the eigenvalues $\xi_1 < \cdots
< \xi_\tau$. Using this formulation, the log prior becomes (up to a constant)
\[
-\frac{\lambda}{2} \beta^\top L_w \beta =
-\frac{\lambda}{2} \theta^\top \Phi^\top_{1:\tau} \Phi \Xi \Phi^\top
\Phi_{1:\tau} \theta =
-\frac{\lambda}{2} \theta^\top \text{Diag}_{i=1,\ldots,\tau}\{\xi_i\} \theta,
\]
or equivalently, $\theta \sim N(0, \lambda^{-1}
\text{Diag}_{i=1,\ldots,\tau}\{\xi_i^{-1}\}).$ This results in the MAP
\begin{equation}
\label{eq:intercept}
\hat{\theta} = \argmin_{\theta} \Big\{
\mathcal{D}\big(Y, g^{-1}(\Phi_{1:\tau} \theta)\big) +
\lambda \theta^\top \text{Diag}_{i=1,\ldots,\tau}\{\xi_i\} \theta \Big\}.
\end{equation}
In Section~\ref{sec:priorinference} we discuss guidelines for specifying basis
rank $\tau$ and roughness penalty $\lambda$.

\subsection{Extending the intercept model}
We want our model to include vertex indexed covariate information leading to
better predictive power and further understanding of the vertex attribute
process.  That is, given $p$ predictors, we wish to regress on
\[
\eta_v = g(\Exp[Y_v]) = \beta_0(v) + x_{1v}\beta_1(v) + \cdots +
x_{pv}\beta_p(v).
\]
We perform the same basis expansion described in the intercept model on
each coefficient, using the first $\tau_j$ eigenvectors of $L$, yielding
\[
\eta_v = \sum_{j = 0}^p x_{jv} \beta_j(v) =
\sum_{j = 0}^p x_{jv} \phi_{\tau v}^\top \theta_j
\]
that is, with $\beta_j(v) = \phi_{\tau v}^\top \theta_j$ where $\phi_{\tau v}$
is the $v$-th row in $\Phi_{1:\tau_j}$ and we identify $x_{0v} = 1$ for the
intercept. We now let
\[
D_X = [\Phi_{1:\tau_o} \;
\text{Diag}_{v \in V}\{x_{1v}\} \Phi_{1:\tau_1} \; \cdots \;
\text{Diag}_{v \in V}\{x_{pv}\} \Phi_{1:\tau_p}]
\]
and $\theta = [\theta_0 \; \theta_1 \; \cdots\; \theta_p]$, allowing us to
write the linear predictor as $\eta = D_X\theta.$  Here, $D_X$ is composed of
block columns, stacked horizontally, while $\theta$ is made up of the
vertically stacked $\theta_j$.  We choose to smooth $\eta$ over the network,
resulting in predictions for the vertex attribute that vary smoothly over the
topology of $G$. This more general specification results in the prior
$\theta \sim N(0, \lambda^{-1} {(D_X^\top L_w D_X)}^{-})$ for the basis
expansion coefficients. Moreover, it extends the posterior estimate
in~\eqref{eq:intercept} to accommodate the roughness penalty
$\lambda \eta^\top L_w \eta = \lambda \theta^\top D_X^\top L_w D_X \theta$:
\begin{equation}
\label{eq:map}
\hat{\theta} = \underset{\theta}{\text{argmin}} \, \Big\{
\mathcal{D}\big(Y, g^{-1}(D_X \theta)\big) +
\lambda \theta^\top D_X^\top L_w D_X \theta \Big\}.
\end{equation}

\subsection{Incorporating abrupt changes}
We next add flexibility to the model, allowing it to detect abrupt changes in
the vertex attribute over the topology of the graph, by introducing a
network-indexed latent binary variable $Z$,
\begin{equation}
Z_v \given \gamma \ind \text{\sf Bern}\Big[
      \text{logit}^{-1}\big(U_v^\top\gamma(v) \big)\Big].
\label{eq:latent}
\end{equation}
Here, a set of predictor attributes $U$ and network-smoothed effects $\gamma$
determine the odds of $v$ belonging to a normal or changed state.
In the context of modeling residential burglary, this variable allows us to
discriminate between two crime rates: if $Z_v = 0$ the vertex is considered to
be in a crime hot zone described by $D_x(v)^\top\theta$. If $Z_v = 1$ the
vertex lies in an area of the city with a ``background'' crime rate,
$X(v)^\top\zeta$ where $\zeta$ is constant across the network.  The latent
effects $\gamma$ assume a basis expansion in a similar manner as the main
effects $\beta$:
$\gamma_j(v) = \phi_{\tau v}^\top \omega_j$, and so the linear effects are
$U_v^\top \gamma(v) = {D_U(v)}^\top \omega$, with $D_U$ defined similarly to
$D_X$.  Together we have
\begin{equation}
\begin{split}
Y_v \given \zeta, \theta, Z_v & \ind \text{\sf Po}\Big[
    \text{exp}\big(Z_v X(v)^\top \zeta + (1-Z_v)({D_X(v)}^\top \theta)\big)\Big] \\
Z_v \given \omega & \ind \text{\sf Bern}\Big[
    \text{logit}^{-1}\big( {D_U(v)}^\top \omega \big) \Big] \\
\theta & \sim N\Big(0, \lambda_\theta^{-1} {(D_X^\top L_w D_X)}^{-}\Big) \\
\omega & \sim N\Big(0, \lambda_\omega^{-1} {(D_U^\top L_w D_U)}^{-}\Big).
\end{split}
\label{eq:model}
\end{equation}

\subsection{Prior and related work}
The single-intercept model of Eq.~\eqref{eq:intercept} is akin to kernel based regression
methods used to operate on discrete input spaces, such as
graphs~\citep[e.g.][Section 8.4]{kolaczyk2009}.
The graph Laplacian is often used in the formation of kernels when
the goal is to approximate data on a graph; see~\cite{smola2003} for a
number of examples of kernels defined via the Laplacian.  Kernel methods
employ a penalized regression strategy, as in Eq.~\eqref{eq:intercept}, where
the predictor variables are derived from the kernel~\citep{kolaczyk2014}.
Similarly, \citet{belkin2004} consider the problem of labeling a partially
labeled graph through regularization algorithms using a smoothing matrix, such
as the Laplacian, and discuss theoretical guarantees for the generalization
error of the presented regularization framework.

While these model formulations capture information in the network topology,
they do not allow us to easily incorporate pertinent covariates.  Kernel
methods have been extended to include information from multiple kernel
functions, each arising from a different data source~\citep{lanckriet2004}.
In this case, the problem is often redefined as determining an optimal set of
weights used to merge the various kernel matrices~\citep{kolaczyk2009}.
However, these methods often lack interpretability and suffer from
computational issues.  Research has also been performed on variable selection
for graph-structure covariates~\citep{li2008}. The proposed procedure
involves a smoothness penalty on the coefficients derived from the Laplacian,
however, in this particular application it is the predictor variables that
represent the vertices in a graph, and the presence or absence of an edge
identifies correlated features. The question of interest revolves around
identifying grouping effects for predictors that are linked in the network.
While the machinery employed is similar to that previously discussed, the
question of interest is essentially different.

With the increased popularity and availability of network data, developing a
framework for regression models specific to network indexed data has become a
focus of recent research. ~\citet{levina2016} discuss network prediction
models that incorporate network cohesion, the idea that linked nodes act
similarly, and node covariates.  They develop the theoretical properties of
their estimator and demonstrate its advantage over regressions that ignore
network information.  Similarly to Li et al.,~\nocite{levina2016}our model
focuses on interpretability and generalization; learning about the network
and the vertex attribute of interest by examining the covariate values and
introducing a flexible framework adaptable to a variety of GLM settings.
However, our method differs in that the coefficients are designed to vary over
the network, addressing the non-homogeneity of covariate effects, and allowing
for an interpretation of how a covariate's influence on the attribute process
of interest changes across the network.  Furthermore, our hierarchical
structure allows for both smooth and abrupt changes in the process rate.  These
are critical extensions in modeling residential burglary, as dynamic patterns
may exist between the residential burglary occurrences and the predictor
attributes that depend on the local structure of the city.

There is also a growing body of literature that utilizes network analysis to
examine the relationship between road structure and residential
burglary~\citep{forecasting, davies,frith}.  Much of this work utilizes
descriptive statistics calculated from the network, such as betweenness or
closeness, and incorporates these metrics into a model as predictor variables.
The structure of the network beyond these statistics is not considered,
however, the cumulative results provide evidence that street network properties
are a significant predictor of burglary.   This illustrates the potential of
leveraging network connectivity to better understand patterns of residential
burglary.   Additionally, there are a number of machine learning algorithms and
software programs being utilized to perform real time predictive analysis of
crime in urban areas~\citep{predictive}.  Our methodology is not intended to
compete with these big data approaches, but it can provide valuable information
regarding current criminological theories.  To the best of our knowledge, ours
is the first implementation of network regularized regression at the
street-segment level for modeling occurrences of urban crime.   Furthermore the
construction of our residential network is a novel contribution to the emerging
research area of crime modeling via network analysis.

\section{Prior Elicitation}
\label{sec:priorinference}
Fitting the proposed model to data requires eliciting
hyper-prior parameters. In particular, we need to select three main sets of
parameters:

\begin{description}
\item[Laplacian weights $w$.] The weights defining the Laplacian $L_w$ measure
similarity between nodes.  In our application, an urban street network, we have
a natural measure of distance: the travel distance between the midpoints of two
street segments.   A network range parameter is introduced to calibrate the
transformation from distance to similarity.

\item[Basis ranks $\tau_{cj}$ and $ \tau_{bj}$.] The cardinalities of the
network basis expansions (differentiated between the count and Bernoulli
levels) for each predictor control how long-range, global network effects in
$D_X$ and $D_U$ are captured. They are elicited based on a modified model using
spike-and-slab variable selection~\citep{george1993}.

\item[Roughness penalties $\lambda_\omega$ and $\lambda_\theta$.]
These prior precision scale parameters control the amount of regularization or
smoothing of coefficients $\gamma$ and $\beta$, respectively. They are
elicited based on a criterion to minimize prediction error, along with a
hyper-parameter controlling the variance between spike and slab when selecting
the ranks.
\end{description}

\subsection{Eliciting weights}
A natural similarity measure between adjacent street segments is an inverse of
street distance. We employ an exponential decay function~\citep{banerjee2014}
to define weights, $w_{ij} \propto \exp\{-d(i, j)/\psi\}$ where $d(i,j)$ is the
travel distance between the midpoint of street segment $i$ and the midpoint of
street segment $j$ and $\psi$ is a range parameter.  We set $\max\{w_{ij}\} =
1$ to avoid identifiability issues with the roughness penalties (see
Section~\ref{sec:casestudy} for details).
%As a
%guideline, we suggest defining $\psi$ such that the similarity weight
%distribution is not too peaked (see sections~\ref{sec:casestudy} for details).
%For instance, as outlined in
%Section~\ref{sec:casestudy}, we define $\psi$ such that the median distance
%maps to 80\% similarity. This pragmatic approach allows the model to
%effectively differentiate between intersections that are close together and
%far apart, ensuring that the range of distances in the city network
%corresponds to an appropriate range of weights.

%TODO: Eric did not like above reasoning in dissertation

\subsection{Controlling basis rank expansions}
\label{sec:estimatingk}
Here we describe the rank selection at the count level of our model.  The
process at the Bernoulli level is analogous, and we drop the count-level
subscript to ease notation.  Now, we would like each $\tau_j$ to be large
enough so that the combination of the $\tau_j$ Laplacian eigenvectors reflects
characteristics of our attribute process while keeping the computational
expense of the model in check~\citep{ramsay2005}.
To this end, we modify the model by setting a hyper-prior on the basis
ranks $\tau_j$ and performing Bayesian variable selection via a spike-and-slab
prior~\citep{george1993} on the basis expansion coefficients.  We adopt a
simpler model structure,
\begin{equation}
\label{eq:ssprior}
\begin{split}
Y_v \given  \theta  & \ind \text{\sf Po}\Big(\text{exp} \big({D_X(v)}^\top \theta\big)\Big) \\
\theta_j \given  \tau_j & \ind N\Big(0,
\lambda^{-1} M_{\tau_j}^{1/2} {(D_{X_j}^\top L_w
D_{X_j})}^{-}M_{\tau_j}^{1/2}\Big)
\end{split}
\end{equation}
where, with $K \leq |V|$ the maximum basis rank, $I$ the indicator function,
and $D_{X_j} = \text{Diag}\{X_j\} \Phi_{1:K}$, we have
$M_{\tau_j} = \text{Diag}_{i=1,\ldots,K} \{I(i > \tau_j)V_0 +
I(i \leq \tau_j)\}$. Here, $0 < V_0 \ll 1$ is a small value, set to 0 in our
application, distinguishing between the variance of the spike and slab
components.  Thus, \eqref{eq:ssprior} is the model induced when the first
$\tau_j$ components of $\theta_j$ are assigned to the slab, and the remainder
to the spike.

For the hyper-prior we set
\begin{equation}
\label{eq:tau}
\Pr(\tau_j) =
  \begin{cases}
    1-\alpha_0, & \tau_j = 0 \\
    \alpha_0 (1-\alpha_1), & \tau_j = 1 \\
    \alpha_0\alpha_1 \rho^{\tau_j - 1} / \sum_{k=2}^{K}\rho^{k-1}, &
    \tau_j = 2, \ldots, K.
  \end{cases}
\end{equation}
Hyper-parameters $\alpha_0 = \Pr(\tau_j > 0)$ and $\alpha_1 = \Pr(\tau_j > 1
\given \tau_j > 0)$ control the prior probability of predictor $X_j$ being
selected and having a network basis expansion, respectively. Thus, parameters
$\alpha_0$ and $\alpha_1$ can be elicited directly based on expert opinion of
the odds of a predictor being included in the model and, if that is the case,
of varying in the network.
Since $\rho = \Pr(\tau_j = i) / \Pr(\tau_j = i - 1)$ for $i = 2, \ldots, K$,
this parameter controls, in effect, the cardinality of the expansion. To
specify $\rho$, we recommend first finding a representative value for
$\bar{\tau} = \Exp[\tau_j]$ and then, given $\alpha_0$ and $\alpha_1$, solving
for $\rho$; see Appendix~\ref{ssec:rho} for details. A reasonable choice for
$\bar{\tau}$ is the smallest value such that
$\sum_{k=2}^{\bar{\tau}} \xi_k^{-1} / \sum_{k=2}^K \xi_k^{-1}$ is
bounded by a large value close to $1$, e.g. $0.9$, similar to the usual
variance explained or scree plot methods used to retain components in
principal component analysis.   Recall, the eigenvectors are ordered by the
eigenvalues, where $\xi_1 < \cdots< \xi_{\tau_j}$, and the structure of the
network may be governed by a small percentage of eigenvectors (similar to how
singular value decomposition can be used to obtain a smaller rank
representation of a matrix).

To select the basis ranks, we employ an ECM algorithm~\citep{mengrubin93}
for model~\eqref{eq:ssprior} where we cycle over each $j$-predictor to
infer conditional posterior modes for $\theta_j$ with $\tau_j$ as a latent
variable~\citep{emvs}. Rather than explore all
possible $2^K$ subsets of eigenvectors of the weighted Laplacian, as usual in
a variable selection setting, we traverse the eigenvectors sequentially to
determine $\tau_j$.  Focusing on the $j$-th predictor and conditional on all
the other predictors, for the E-step we need, at the $t$-th iteration,
$\nu_{ji}^{(t)} \doteq \Pr(\tau_j < i \given Y, \theta_j^{(t)})$,
$i = 1, \ldots, K$. For the CM-step, we set $\theta_j^{(t+1)}$ by Poisson
regressing $Y$ on $D_{X_j}$ with prior precision
$\lambda (T(\nu_j^{(t)}) \circ (D_{X_j}^\top L_w D_{X_j}))$
where $D_{X_j} = [\text{Diag}\{X_j\} \Phi_{1:K}]$, operator $\circ$ is the
Hadamard (element-wise) product,
\[
T(\nu) = \Big[1 - \nu_{\max\{i, k\}} +
\big(\nu_{\max\{i, k\}} - \nu_{\min\{i, k\}}\big)V_0^{-\frac{1}{2}} +
\nu_{\min\{i, k\}} V_0^{-1}\Big]_{i,k = 1, \ldots, K},
\]
and an offset composed of the information from the remaining
predictors in the model. After convergence, we select $\tau_j$ via
\emph{sequential centroid estimation}, which tends to be more robust than
the usual conditional posterior mode and, in effect, picks the minimum number
of eigenvectors such that the cumulative posterior is less than some
threshold. This procedure is repeated for each one of the predictors
in turn, continuously updating prior precisions and offsets, until $\tau$ does
not change between consecutive cycles.  See Appendices~\ref{ssec:tauj}
and~\ref{ssec:ecm} for detailed presentations of the ECM algorithm and the
sequential centroid estimator.

\subsection{Selecting  $\lambda_\theta$, $\lambda_\omega$}
\label{ssec:penalty}
We use a leave-one-out cross validation PRESS statistic defined
on the working responses in the final step of the iterative reweighted least
squares (IRLS) algorithm, the usual computational routine used to fit
generalized linear models~\citep{mccullagh1989}. We call this the LOOP
(leave-one-out-proxy) statistic,
\[
\text{LOOP} = \sum_{v \in V}
\frac{{(Y_v-\hat{\mu}_{(v),v})}^2}{V(\hat{\mu}_{(v),v})}
\approx \sum_{v \in V} \frac{r_v^2}{1-h_{v}},
\]
where $\hat{\mu}_{(v),v}$ is the mean at $v$ fitted without the $v$-th
observation, $V(\mu)$ is the variance function ($V(\mu) = \mu$ for the Poisson model),
$r_v = (Y_v-\hat{\mu}_v) / \sqrt{V(\hat{\mu}_v)}$ is the $v$-th Pearson
residual, and $h_v$ is the leverage at $v$.
%While a closer approximation is
%provided by~\citet{williams1987}, our formulation is more computationally
%convenient.

We define $D_X$ and $D_U$ using the ranks found by the procedure in
Section~\ref{sec:estimatingk} and then minimize the LOOP statistic to determine
$\lambda_\theta$ and $\lambda_\omega$ respectively.  This process can be viewed
as an empirical Bayes procedure where we maximize prediction accuracy rather
than posterior density.

\section{Model Inference}
\label{sec:modelinference}
%After specifying hyper-parameters and determining the dimensions of $D_X$ and $D_U$, we
%are in position to estimate model parameters. To this end,
%
To estimate the model parameters, we sample from
the joint posterior $\Pr(Z, \theta, \omega, \zeta \given Y)$ using Gibbs
sampling. We iterate sampling from the conditional distributions
\begin{equation}
[Z \given \omega, \zeta, \theta, Y],
\quad
[\zeta \given Z, \omega, \theta, Y],
\quad
[\theta \given Z, \omega, \zeta, Y],
\quad
[\omega \given Z, \omega, \zeta, Y],
\label{eq:conditional}
\end{equation}
until assessed convergence. Sampling Z is straightforward; we have
\[
\Pr(Z \given \omega, \zeta, \theta, Y) \propto
\Pr(Y \given \zeta, \theta, Z) \Pr(Z \given \omega) =
\prod_v \Pr(Y_v \given \zeta, \theta, Z_v) \Pr(Z_v \given \omega)
\]
and so, referencing ~\eqref{eq:model}, we have
$Z_v \given \omega, \zeta, \theta, Y_v \ind \text{\sf
Bern}\big(\gamma_v(\omega, \zeta, \theta)\big)$
with
\begin{equation}
\text{logit}\gamma_v(\omega, \zeta, \theta) \doteq
D_U(v)^\top \omega + Y_v X(v)\zeta - \exp (X(v)\zeta) -
\big(Y_v D_X(v)^\top \theta - \exp(D_X(v)^\top\theta) \big).
\label{eq:gamma}
\end{equation}
To sample $\zeta$, $\theta$, and $\omega$ we first note that the likelihood
in~\eqref{eq:model} can be rewritten to partition the data $Y$ according to
the latent indicators as
\[
Y_v \given Z_v = 1, \zeta \ind \text{ \sf Po}(\exp (X(v)^\top\zeta))
\quad \text{and} \quad
Y_v \given Z_v = 0, \theta \ind
\text{\sf Po}\big(\exp({D_X(v)}^\top\theta)\big).
\]
Sampling in the three last conditional steps in~\eqref{eq:conditional} is then
equivalent to sampling coefficients from the posterior of a Poisson log-linear
model, in the case of $\zeta$ and $\theta$, and a logistic model, in the case
of $\omega$, with normal priors as stated in~\eqref{eq:model}. Each
conditional step is a Metropolis-within-Gibbs step, using a one-step
Riemannian manifold Hamiltonian Monte Carlo proposal, also known as a
manifold Metropolis adjusted Langevin algorithm \citep[MMALA; see][]{sampling2}.
We describe this process and outline the proposal density and acceptance
probability in Appendix~\ref{sampling}.

Due to the potentially large scale of data in common applications, the above
sampling scheme may be slow to converge.  We thus suggest finding
initial values for the sampler via an EM algorithm with $Z$ as a latent
variable; see details in~\ref{hotzone}. For the E-step we compute, at
the $t$-th iteration,
$\pi_v^{(t)} = \Exp[Z_v \given \omega^{(t)}, \zeta^{(t)}, \theta^{(t)}, Y_v] =
\gamma_v(\omega^{(t)}, \zeta^{(t)}, \theta^{(t)})$, with $\gamma_v$ as
in~\eqref{eq:gamma}. The three M-steps are then as follows:
\begin{description}
\item[M-step for $\zeta$:] set $\zeta^{(t+1)}$ by quasi-Poisson
regressing $\pi^{(t)} Y \sim X$ with offset $\log\pi^{(t)}$;
\item[M-step for $\theta$:] set $\theta^{(t+1)}$ by quasi-Poisson
regressing $(1-\pi^{(t)})Y \sim D_X$ with offset $\log(1-\pi^{(t)})$ and
$\lambda_\theta D_X^\top L_{w(\psi)} D_X$ as prior precision;
\item[M-step for $\omega$:] set $\omega^{(t+1)}$ by quasi-binomial
regressing $\pi^{(t)} \sim D_U$ with
$\lambda_\omega D_U^\top L_{w(\psi)} D_U$ as prior precision.
\end{description}

\section{Data Analysis and Results}
\label{sec:analysis}
We now turn to our application and conduct two studies around
residential burglary occurrences in Boston, MA: a simulation study and a more
detailed case study. The data, provided by the city of Boston and available to
the public~\citep{bostondata,opendata} contain information on the 5,741
instances of residential burglary occurring between June 2015 and December 2019
in central Boston.  The covariate data, as described below, was constructed
from a variety of publicly available shape files.   The code used to perform
the analyses, with links to shape files, is available in the supplementary
materials.

\begin{figure*}
\centering
\includegraphics[scale = .35]{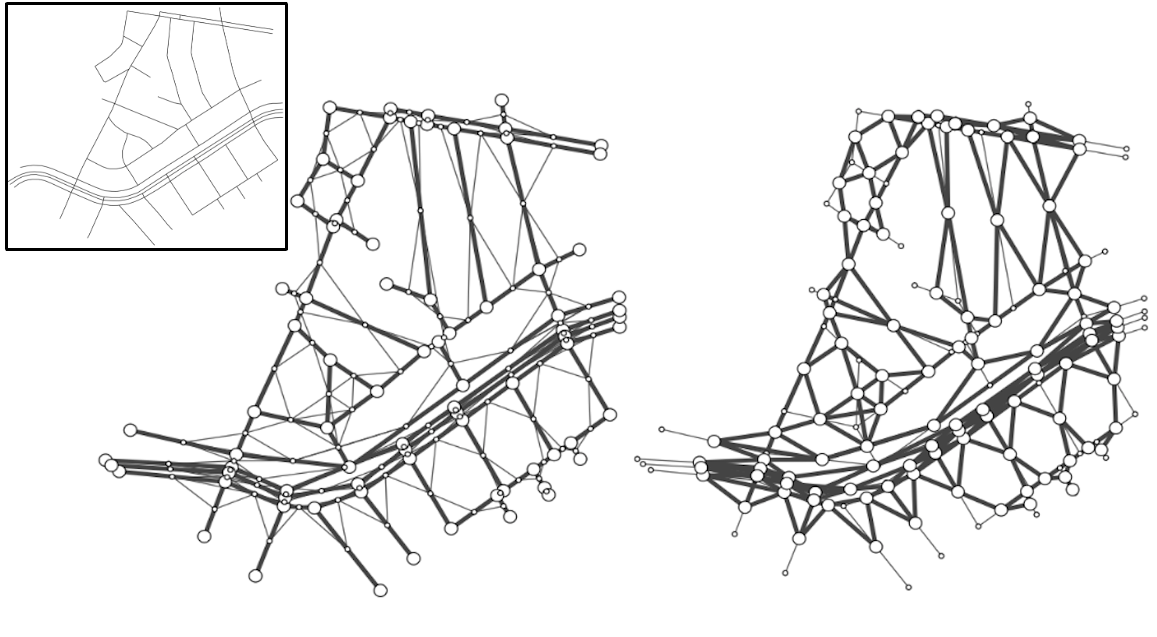}
\caption{Comparison of primal (left) and dual (right) network representations.
To ease comparison, each network is superimposed on the other (thin edges and
small nodes).   The black box in the upper left is the map visualization of the
area.}
\label{fig:primaldual}
\end{figure*}

The first step in applying our modeling procedure is to clearly define our
network of interest.   A city can be intuitively represented as a network by
designating street intersections as nodes and street segments as edges, as
visualized for a small area of Boston in the left panel of
Figure~\ref{fig:primaldual} (this is often referred to as the primal
representation).   However, given that residential burglary predominantly
occurs along street segments rather than at intersections, utilizing street
segments as the main unit of analysis, i.e.  the vertices,  yields a more
representative network. Moreover, given that the exact location of each crime is sensitive, recording
it in aggregated form by street segment guarantees some level of
confidentiality.  This formation is pictured in the right panel of
Figure~\ref{fig:primaldual} and is often referred to as the dual
representation~\citep{dual}.  In graph theory, the dual network is also known
as the line graph of the primal network.  We note that while the edges are
visualized as straight lines, the weight of each edge is the street distance
from the midpoints of the street segments (not the Euclidean distances).
This better captures the connectivity of the city through the perspective of a
potential criminal.

\begin{figure*}
\centering
\includegraphics[scale = .4]{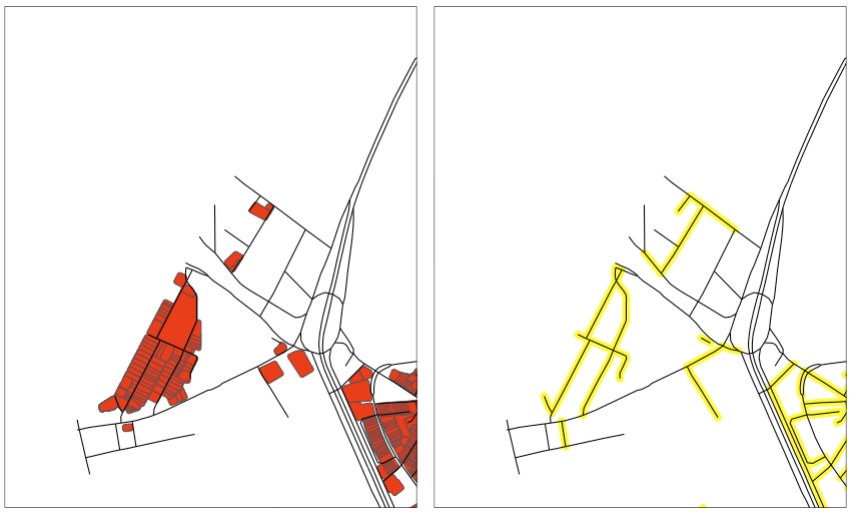}
\caption{Illustration of residential parcels in red (left) and street segments
remaining in the vertex set in yellow (right) on a small region in Boston, MA.}
\label{fig:marginalization}
\end{figure*}

Although we have covariate and response data for each street segment in the
city of Boston, only those segments in residential areas can experience the
response of interest, occurrences of residential burglary.  Thus, we want to
restrict the vertex set to street segments that abut residential parcels while
not losing the connectivity information in the entire street network.  We
accomplish this through the Kron reduction process~\citep{kron}, which
maintains effective resistances (graph-theoretic distances between nodes) while
reducing the node set.  Specifically, given a set of $v$ vertices to be
marginalized out in a weighted graph, we need to compute the Laplacian of
$G[-v]$ in order to fit our model.
%
%\begin{figure*}
%\centering
%\includegraphics[scale=.25]{schur.png}
%\caption{Marginalization process on a toy network, using the Schur complement of the Laplacian matrix, that maintains connectivity between the node set.}
%\label{fig:schur}
%\end{figure*}
%
Partitioning the street segments in our network into residential, $R$, and
non-residential, $N$, segments, we can similarly partition the weighted
Laplacian $L$ of $G$ into blocks $L_{RR}$, $L_{RN} = L_{NR}^\top$, and
$L_{NN}$. We then find the weighted Laplacian of $G[-N]$ to be
$L_{RR} - L_{RN}L_{NN}^{-1}L_{NR}$, the Schur complement of $N$ in $L$.
%
%An example of this process is provided in Figure~\ref{fig:schur}, where node 4
%has been marginalized out from the toy network and similarity measurements, or
%weights, between nodes 1-2, 1-3, and 2-3 are maintained.
Our resulting \textit{residential} network of Boston is composed of 12,763
street segments (as compared to 15,078).  Figure~\ref{fig:marginalization}
illustrates the process on a small section of Boston, where the left image
displays the residential parcels in the city, and the right displays the street
map.  Street segments highlighted in yellow remain in the vertex set, but those
in gray are marginalized out.  As expected, all street segments marginalized
out from the network saw zero occurrences of residential burglary in the time
frame of the data.

The street segment attributes gathered and included in our final
model are:
\begin{description}
\item[Population:] population size, based on census tracts from the US Census Bureau, 2010
\item[Gross tax:] property tax surrounding each street segment.  To calculate,
we construct a buffer around each street segment, capturing the residential
partials abutting the street, and then sum the gross tax from each parcel.  The
results are visualized in Figure~\ref{fig:bostondescriptive}.
\item[Distance to nearest police station:] street distance from the center of
each segment to the nearest police station.
\item[University/college housing indicator:] binary indicator set equal to one
if college/university housing is located on the street segment.
\end{description}

The population effect is assumed constant across the network, that is, we
consider it as an exposure variable, while the other covariate effects are
subject to the results of the basis rank selection. Our choice of analyzing
the aforementioned covariates was driven by established crime
theory~\citep{bernasco2009} and publicly available data. The model could
easily be extended to include additional covariates of interest.
\begin{figure*}
\centering
\includegraphics[scale=.5]{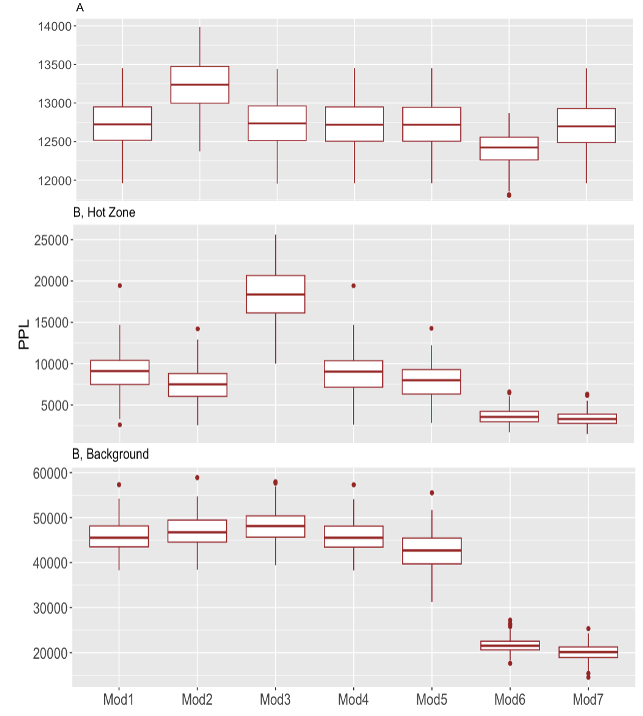}
\caption{We compare posterior predictive loss (PPL) for two simulations.  A:
Simple data generating process, B: Complex data generating process.  For
simulation B, we present PPL for street segments in and out of hot zones
according to the simulation scheme.}
\label{fig:simstudy}
\end{figure*}
\subsection{Simulation study}
In order to assess the performance of our model (\textbf{Mod7},
in~\eqref{eq:model}) we compare its output in two different simulation studies
against six competing methods:
\begin{description}
\item[\textbf{Mod1}:] Intercept-only Poisson regression, akin to Laplacian kernel
regression, $Y_v \given \theta \ind \text{\sf Po}[\exp(\phi_{\tau v}^\top\theta)]$
and $\theta \sim N(0,
\lambda^{-1} {\text{Diag}_{i=1, \ldots, \tau}\{\xi_i\}}^{-})$.
\item[\textbf{Mod2}:] Intercept-only Poisson regression, akin to kernel
regression, utilizing the diffusion kernel of \cite{smola2003}.
\item[\textbf{Mod3}:] Poisson regression using the covariates, but ignoring
the network similarity,
$Y_v \given \theta \ind \text{\sf Po}[\exp(\mathbf{x}_v^\top \beta)]$.
\item[\textbf{Mod4}:]  A Poisson network cohesion model in the flavor of that
proposed by~\cite{levina2016}. In essence, we assume a basis expansion of the
intercept only and set all other basis ranks, $\tau_j$, to one.

\item[\textbf{Mod5}:] Poisson regression defining $D_X$ and smoothing the
linear predictor, but ignoring the abrupt changes in the network,
$Y_v \given \theta \ind \text{\sf Po}[\exp({D_X(v)}^\top \theta)]$ and
$\theta \sim N(0, \lambda^{-1} {(D_X^\top L_w D_X)}^{-})$.

\item[\textbf{Mod6}:] Intercept-only Poisson regression, akin to kernel
regression, that takes into account abrupt changes in the network
$Y_v \given \zeta, \theta, Z_v  \ind
\text{\sf Po}[\exp(Z_v \zeta + (1-Z_v)\phi_{\tau v}^\top\theta)]$,
$Z_v \given \omega \ind \text{\sf Bern}\Big[\text{logit}^{-1}\big(
\phi_{\tau v}^\top\omega \big) \Big]$,
$\theta \sim N(0,\lambda^{-1} {\text{Diag}_{i=1, \ldots, \tau}\{\xi_i\}}^{-})$
and
$\omega \sim N(0,\lambda^{-1} {\text{Diag}_{i=1, \ldots, \tau}\{\xi_i\}}^{-})$.
\end{description}
For the simulations, we use the residential street network ($n$ = 1,278)
information for Allston/Brighton (the northwest peninsula neighborhoods of
Boston, pictured in Figure~\ref{fig:Neighborhoods} in the appendix).  This part
of the city is not included in our case study (see discussion below), however,
we make use of the available covariate and network structure for two
simulations:
\begin{itemize}
\item A: We generate random crime counts on each street segment from a Poisson
distribution with constant mean.
\item B: We generate random crime counts on each intersection using the
negative-binomial distribution with mean $\mu = \exp(Z_v \zeta +
(1-Z_v)({D_X(v)}^\top \theta))$ and variance $\mu + \mu^2$.  We chose to sample
from the negative-binomial distribution given that crime data are often
over-dispersed.   For $Z_v, $ we designate three hot zones, using the
breadth-first search algorithm originating at three randomly chosen
intersections and capturing 10\% of the nodes~\citep{cormen2001}.  To construct
$D_X$ we set a universal $\tau_j$, and sample $\theta$ from the prior
distribution given in~\eqref{eq:model}.  Lastly, we set $\zeta$, a background
crime rate, equal to -2.5 which was found to create a range of crime counts
similar to that of the observed data.
\end{itemize}

We complete the seven regressions of interest and compare their performance
based on posterior predictive loss, a proxy for out of sample prediction.  For
this exercise, we forgo our Gibbs sampling algorithm and use the posterior
modes calculated through the EM-algorithm as coefficient estimates.  For
\textbf{Mod1,} \textbf{Mod2}, \textbf{Mod4} and \textbf{Mod5,} we choose $\tau$
based on the LOOP.  For \textbf{Mod6} and \textbf{Mod7} we define $\tau_j$ and
$\lambda$ using the procedure described in Section~\ref{sec:estimatingk}.

As shown in Figure~\ref{fig:simstudy}, \textbf{Mod7} does not degrade in
performance due to increased model complexity.  That is, the model has
protection against overparameterization due to the rank selection process.
In~139 out of the~200 simulations, the $\tau_j$ value for each of the $j$
predictors was~1, indicating the process determined that a basis expansion was
not informative.

\textbf{Mod7} outperforms the competing methods for intersections in and out of
the three hot zones. Furthermore, \textbf{Mod5} shows improvement over
\textbf{Mod4}, indicating that \textbf{Mod5} (and \textbf{Mod7}) explains
additional variation in crime counts by allowing the coefficient effects to
vary smoothly over the network.   \textbf{Mod6} and \textbf{Mod7} illustrate
the effectiveness of the hotzone indicator to differentiate between the two
different crime rates.
\begin{figure*}
\centerline{\includegraphics[scale = .35]{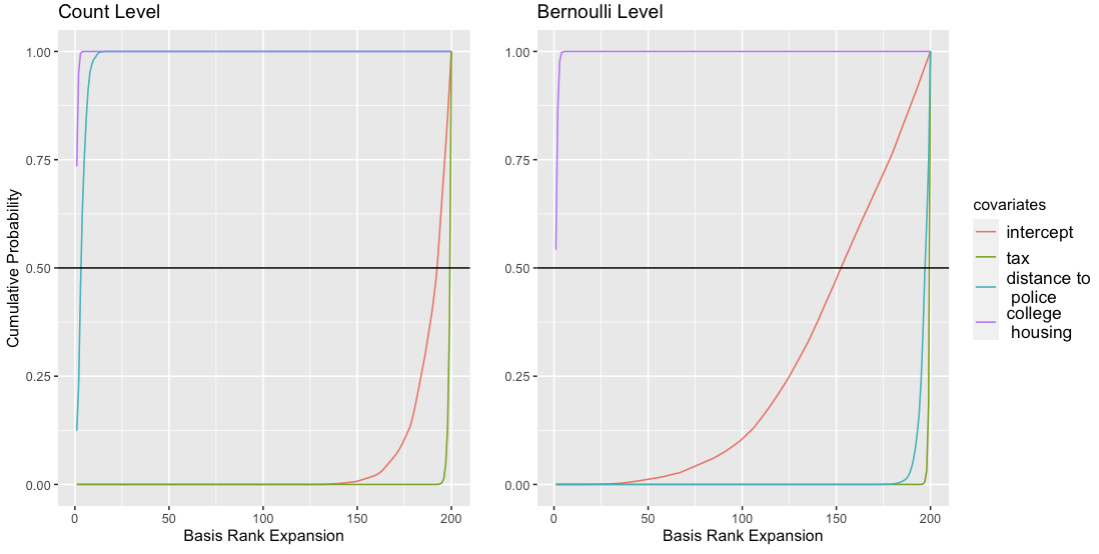}}
\caption{Choosing $\tau$, the rank of the coefficient effects, based on a
threshold of 0.5.}
\label{fig:expansion}
\end{figure*}

\subsection{Case study: Boston, Massachusetts}
\label{sec:casestudy}
We now analyze the entire metropolitan region of Boston, excluding
Allston/Brighton and East Boston (see Figure~\ref{fig:Neighborhoods}).  These
parts of Boston are separated from Boston proper:  East Boston can only be
accessed via ferry boat or tunnel and Allston's southern border town
(Brookline) is not part of Boston.  Because we don't have access to network and
crime reporting data for Brookline,  the network depiction of the city suggests
a disconnection of Allston from Boston that is misleading.  Numerous pathways
traverse Brookline, linking Allston to Boston, and the absence of this
information results in an inaccurate representation of Allston's connectivity
with the broader city.   That being said, one could perform the analysis
separately on these two neighborhoods.

Following the guidelines in Section~\ref{sec:priorinference}, the network range
parameter, $\phi,$ is set to 0.15, resulting in a median similarity weight of
0.8 in the adjacency matrix. All continuous predictor variables were
log-transformed and scaled. The background crime rate, described by
$X(v)^\top\zeta$ in~\eqref{eq:model}, is based only on population.  The maximum
basis expansion rank was set to 200 (based on exploratory analysis), and,
following Section~\ref{sec:priorinference}, $\alpha_0 \doteq 1, \alpha_1 \doteq
0.99$ and $\sum_{k=2}^{\bar{\tau}} \xi_k^{-1} / \sum_{k=2}^K \xi_k^{-1}$ is
bounded by 0.9 in order to determine $\bar{\tau} = 68.$  These hyperparameters
were chosen to give relatively high probabilities of variable inclusion and
expansion.  For the ECM algorithm in Section~\ref{sec:estimatingk}, $V_o$ is
set to zero, and Figure~\ref{fig:expansion} displays the cumulative posterior
results, $\Pr(\tau_j\given Y, \theta_j)$ for each predictor at both the count
and Bernoulli levels of the model.  We included the same set of predictor
variables in both levels of the model, but this is not required.

As posited in the introduction, the gross tax effect varies over the network
and is best captured via a basis expansion of large rank; conversely, distance
to the nearest police station requires a smaller basis expansion at the Count
level of the model and the effect of college housing appears to be constant
across the network.  The connection between the covariate and its rank gives
some indication of the complexity of the variable's relationship to the process
rate being modeled.  For example, given that most college housing occurs in two
small areas of Boston proper, it makes sense that the effect of this indicator
variable on residential burglary counts does not vary across the city.   Given
the basis expansion ranks, we construct $D_X$ and $D_U$ and, using the LOOP,
determine $\lambda_\theta$ and $\lambda_\omega$.
%FIXME
%The predicted crime counts from model~\eqref{eq:map} define the initial values of $Z_v$, $\theta$, $\omega$, and $\zeta$ in the second EM algorithm. Specifically, for the %vertices in the upper quartile of predicted
%crime occurrences, we set $Z_v$ equal to 0; this subset of intersections is used via
%model~\eqref{eq:map} to find initial estimates of $\theta$ and $\omega$. The remaining points define the initial value for $\zeta$.
\begin{figure*}
\centerline{\includegraphics[scale=.25]{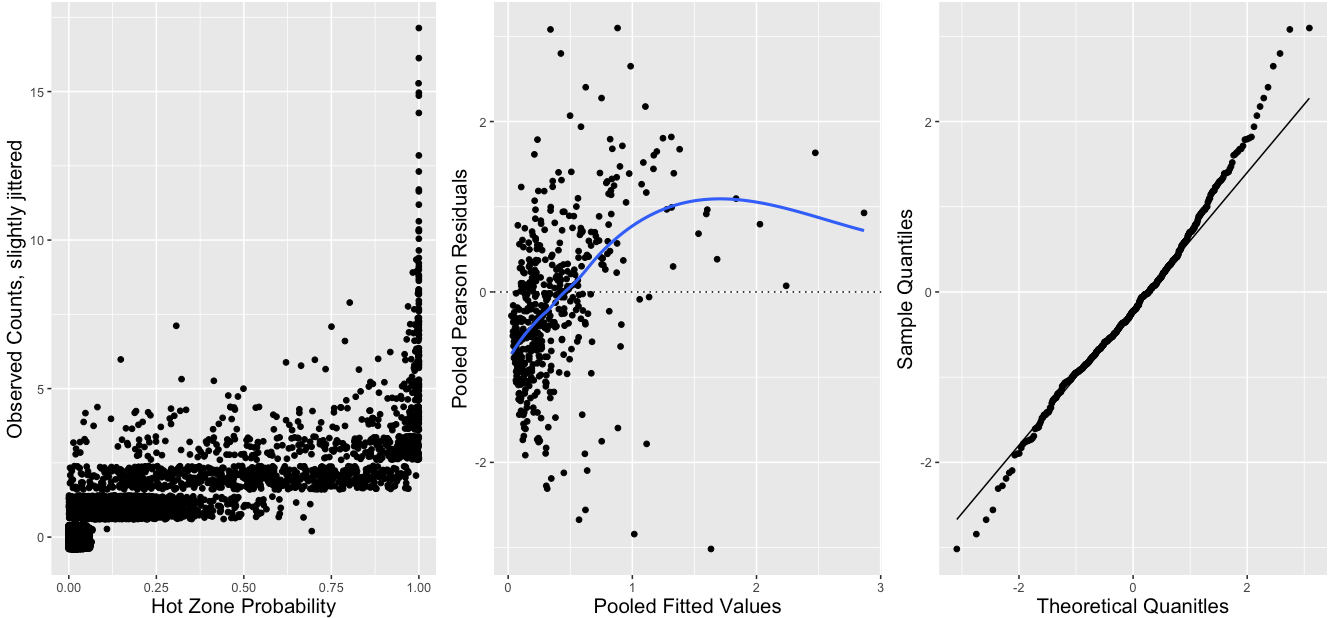}}
\caption{Observed counts versus latent values, $1-\pi_i$ (left), Pooled Pearson
residuals versus fitted values (middle) and normal quantile comparison plot
(right).}
\label{fig:diagnostics}
\end{figure*}
The MAP estimates found using the EM algorithm are used as initial values in
the Gibbs sampler.  Figure~\ref{fig:Gibbs}, included in the appendix, displays
the posterior distributions of a selection of statistics, and are based on 1000
samples, after discarding 20\% for a burn-in period and thinning every 10.  We
ran multiple chains and assessed convergence of the Gibbs sampler by analyzing
trace plots and evaluating rank-based $\hat{R}$ and effective sample
size~\citep[ESS;][]{rhat}.   Of note, the EM algorithm provides reasonable
point estimates and is computationally efficient (we include supporting
evidence of this in Appendix~\ref{samplingresults}).

Diagnostic plots from the fitted model are in Figure~\ref{fig:diagnostics}.  We
see that the hot zone probabilities align well with the observed counts of
residential burglary.  The plot of the binned Pearson residuals (response
residuals divided by the non-constant variance estimate, as typically utilized
when assessing a GLM) versus the fitted values is relatively null given our
dependent variable is discrete.   Furthermore, the normal quantile plot  of the
Pearson residuals does not suggest a strong departure from normality; while
crime data is often over-dispersed or zero-inflated, the creation of our
\textit{residential} network and the hot zone indicator addresses these issues.
However, the outlined methodology, including the EM algorithms, is easily
adapted to other exponential random family distributions.  For example, we
explored utilizing the negative-binomial distribution and found that the
additional precision parameter in the distribution introduces too much
flexibility.  That is, using the Poisson distribution, the distribution of hot
zone probabilities is bi-modal, with peaks near 0 and 1.  Using the
negative-binomial distribution resulted in the distribution having greater mass
towards 0.5, creating ``luke warm'' zones.  This decreased the predicted
crime rates for intersections located in hot zones and underestimated the
occurrences of residential burglary.
\begin{figure*}
\centerline{\includegraphics[scale = .35]{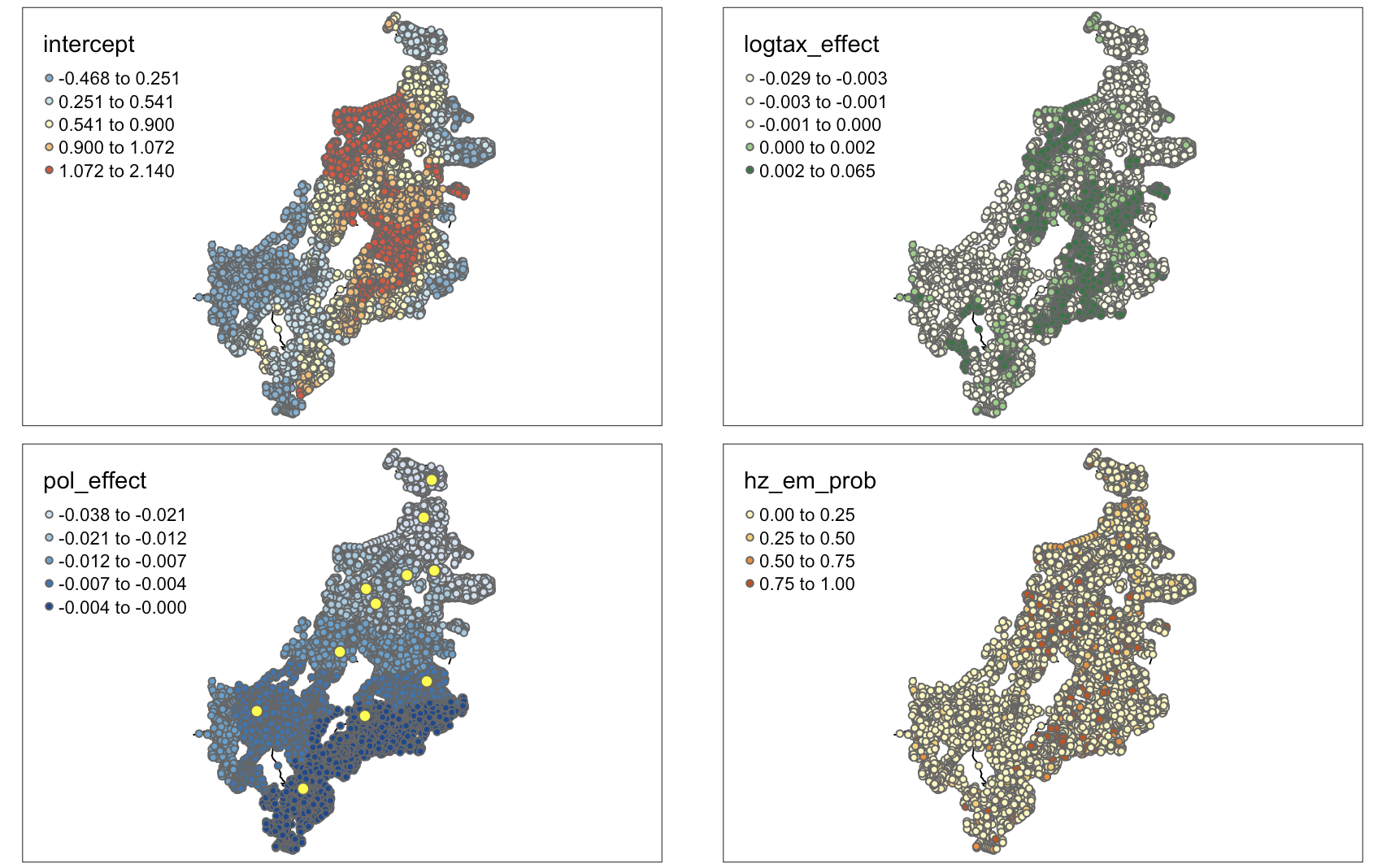}}
\caption{Covariate effects across the city for the count level of the model.
Yellow bubbles on police effect (bottom left) indicate the location of police
departments. The bottom right plot shows the latent values, $1-\pi_i$
describing the probability of each intersection being in an hot zone.}
\label{fig:effects}
\end{figure*}

The interpretational power of our model is of primary interest.
Figure~\ref{fig:effects} summarizes the effects, $\beta(v) =
\phi^{T}_{\tau}{\theta}$, of the intercept, wealth/tax, and distance to nearest
police station.  While the exact values for each street segment are not
displayed, the gradation scales allow us to identify areas of the city where
the effects of each covariate are most pronounced (in this case, the
coefficients have the same interpretation as in a Poisson regression).
Individual estimates and significance levels per street segment could be
evaluated by referencing the posterior distributions.

On the tax effect plot, notice the large number of dark green intersections
down the central corridor of the city (originating in the Fenway/South End
neighborhood down through Mattapan and South Dorchester), identifying areas
where the tax effect on residential burglary is greatest.  It is noteworthy
that this corridor was recently identified as socially
vulnerable~\citep{carbon} and highly susceptible to  gentrification and
displacement~\citep{displacement}.  Boston's published displacement indices are
calculated based on 16 indicators encompassing demographic factors (e.g. race,
housing cost burden, educational attainment),  amenities (e.g. access to public
transportation) and market changes (e.g. rent appreciation, commercial
development).  While the connection between urban displacement and crime levels
is an area of active research with some evidence supporting gentrification
reducing crime~\citep{gentrification}, it could also be that these neighborhood
dynamics result in target rich environments for residential burglary.   More
specifically, residential parcels with higher property taxes may be
particularly susceptible to occurrences of residential burglary in this area of
the city.

\begin{figure*}
\centerline{\includegraphics[scale = .35]{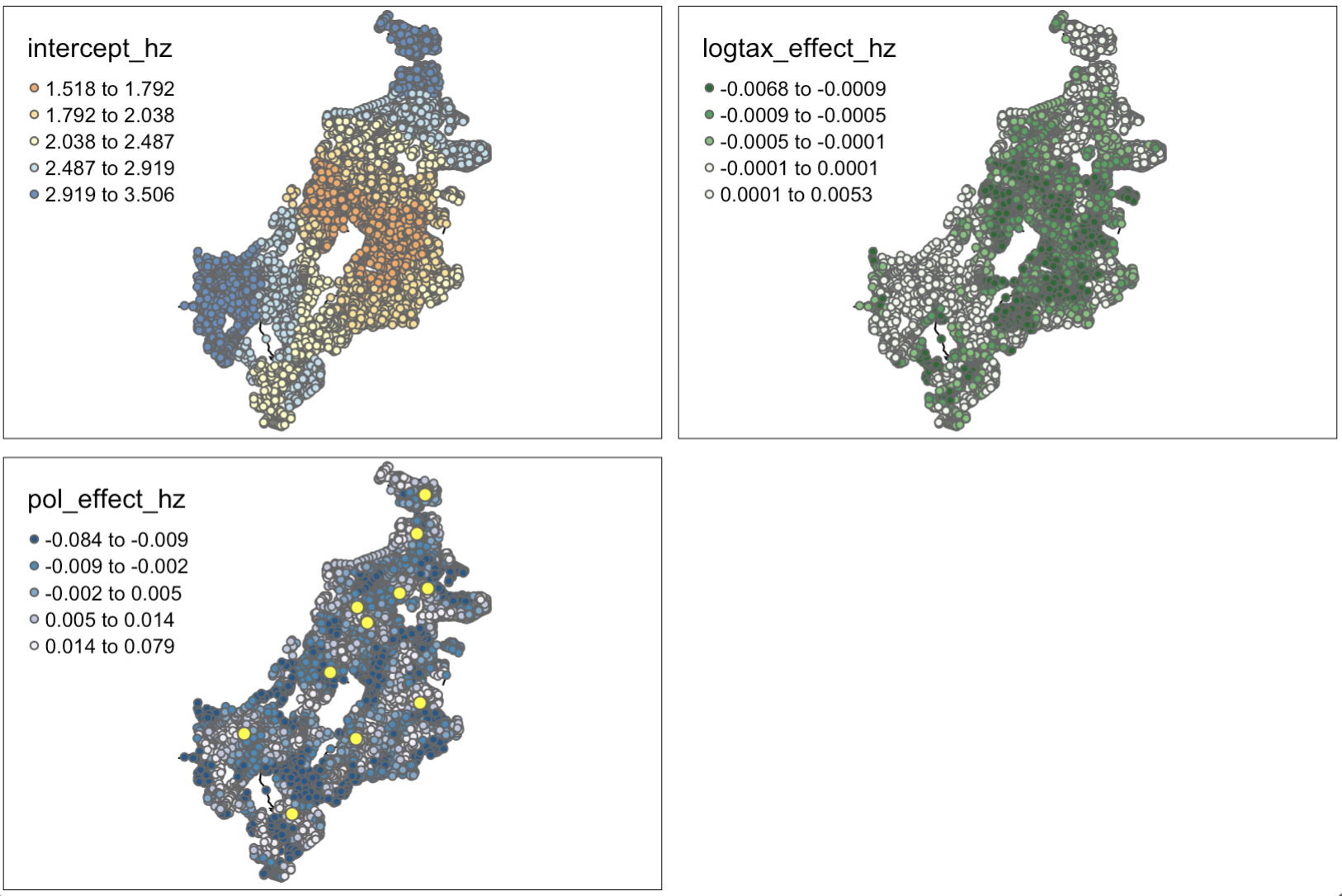}}
\caption{Covariate effects across the city for the hot zone indicator level of
the model. Recall, if $Z_v = 0$ the vertex is considered to be in a crime hot zone, so
negative effects contribute to a larger probability of a street segment being
designated as in a hot zone.}
\label{fig:bineffects}
\end{figure*}

Notably, the police effect at the count level of the model is quite smooth
across the city given its small basis rank expansion.  However,
Figure~\ref{fig:bineffects}, illustrating the effects at the hot zone indicator
level, tells a more nuanced story.   Here we see darker street segments
(identifying the larger covariate effects) between the southern neighborhoods
of the city (Mattapan, Hyde Park, and Roslindale).  There also exists strong
effects on the northern border of the open space, Franklin Park, in the center
of the city (Roxbury).  This identifies areas of Boston where a lack of police
presence may be contributing more greatly to the observed crime than in other
areas. Looking at Figures~\ref{fig:effects} and~\ref{fig:bineffects}
collectively gives evidence that proximity to police contributes more to the
hot zone status of a street segment and, given that status, the wealth effect
is more informative at the Count level.

Additional predictors could be included in the outset of the model formulation
to identify their effects across the city.  This information is beneficial to
residents of Boston, local law enforcement agencies seeking to optimize their
resources, and urban planners responding to the city's evolving needs.

%For example,current initiatives for the city of Boston's planning and developmental agency (BPDA) include the Roxbury Strategic Master Plan, aimed at creating "a high quality physical environment that is attractive, safe, and convenient for residents" ~\citep{bostonplans}.
%
%
\section{Conclusions}
\label{sec:conclusions}
We have presented a method of Bayesian network regularized regression for
modeling vertex attributes that incorporates both relevant covariates and
topological information in its construction.  The resulting model is composed
of node indexed coefficients, regularized using the Laplacian matrix,
producing fitted values that vary smoothly over the network.  This machinery,
including the corresponding variable selection techniques and EM algorithms, is
widely applicable and easily adaptable to a variety of GLM settings.  As
previously mentioned, in general, we could regress a response attribute $Y$ on
network-indexed coefficients,
$Y_v \given \beta \ind \text{\sf F}[g^{-1}(\mathbf{x}_v^\top \beta(v))]$, where
{\sf F} belongs to the exponential family and $g$ is a link.  Furthermore, the
described methods can be modified to include specific characteristics of the
network being analyzed (e.g. hot spots); the flexibility of the introduced
contributions is valuable.

Our network approach to modeling occurrences of residential burglary allows us
to incorporate environmental and structural barriers in the connectivity of the
city.  The similarity calculations are based on a metric of walkability and
naturally lend themselves to describing possible patterns in residential
burglary.  Perhaps most significantly, the model results provide a clear
picture of how factors contributing to the response variable vary across the
network. This is particularly useful when applied to modeling residential
burglary: crime rates are not homogeneous, and understanding why specific
regions or street segments experience high crime is valuable information.

Our analysis has prompted a variety of additional research inquiries that could
extend the current work. For example, while we have chosen to focus on modeling
the spatial variation in residential burglary, many crime prediction algorithms
are concerned with temporal variations or the interplay between time and space.
A relatively straightforward way to extend the current methodology would be to
fit distinct models to consecutive time frames of interest and analyze
variations in the coefficient estimates over time.   This mirrors approaches
aimed at quantifying the efficacy of crime intervention initiatives, where
model outcomes predicting the probability of a crime are compared using data
from before and after the intervention occurs~\citep{prevent}.  A more
ambitious approach could be to develop a Bayesian dynamic model that allows the
coefficients to evolve over both time and space, and for the probability
distributions to up updated as new data becomes available.

\newpage

\appendix
\section{Prior Elicitation}

\subsection{Selecting $\rho$}
\label{ssec:rho}
Given hyper-prior parameters $\alpha_0$ and $\alpha_1$ and mean basis rank
$\bar{\tau}$, we need to solve, according to the prior on $\tau_j$
in~\eqref{eq:tau},
\[
\begin{split}
\Exp[\tau_j]
&= \alpha_0 (1 - \alpha_1) + \alpha_0 \alpha_1
\frac{\sum_{k=2}^K k \rho^{k-1}}{\sum_{k=2}^K \rho^{k-1}} \\
&= \alpha_0 (1 - \alpha_1) + \alpha_0 \alpha_1
\Bigg(1 + \frac{\sum_{k=1}^{K-1} k \rho^k}{\sum_{k=1}^{K-1} \rho^k}\Bigg)
= \bar{\tau}.
\end{split}
\]

Let$S(\rho) \doteq \sum_{k=1}^{K-1} k \rho^k$.  Then
$(1-\rho)S(\rho) = \sum_{k=1}^{K-1} \rho^k - (K-1) \rho^K$ and
$\sum_{k=1}^{K-1} \rho^k = \rho(1 - \rho^{K-1})/(1 - \rho)$, thus,
\[
\frac{\bar{\tau}-\alpha_0}{\alpha_0 \alpha_1} =
\frac{S(\rho)}{\sum_{k=1}^{K-1} \rho^k} =
\frac{1}{1-\rho} - \frac{(K-1)\rho^{K-1}}{1-\rho^{K-1}}.
\]
The solution, constrained to $\rho > 0,$ can be solved numerically using
standard procedures such as Newton's method.

\subsection{Selecting $\tau_j$}
\label{ssec:tauj}
To select $\tau_j$ (we drop the subscript $j$ for the remainder of this
discussion), the rank of the basis expansion for each covariate in the
model, we adopt a \emph{sequential centroid estimator}. Let us first define an
auxiliary variable $\omega(\tau)$ that represents $\tau$ as an indicator
vector: $\omega(\tau)_i = I(i \leq \tau)$, for $i = 1, \ldots, K$. For
instance, $\omega(0) = (0, 0, \ldots, 0)$, $\omega(1) = (1, 0, \ldots, 0)$,
and so on, with $\omega(K) = \mathbf{1}_K$. Note the one-to-one correspondence
between $\tau$ and $\omega$, and thus, while
$\tau \in \mathcal{T} \doteq \{0, \ldots, K\}$,
$\omega$ only takes values in $\Omega \doteq \cup_{j \in \mathcal{T}} \omega(j)$.

Now, given the marginal posteriors $\Pr(\tau \given Y)$ or the EM-conditional
posteriors $\Pr(\tau \given Y, \theta^{(t)})$, which we denote in general by
$\pi_{\tau}$, we define a Bayes estimator $\hat{\tau}$ according to a
generalized Hamming gain $G$ on the $\omega$-map:
\[
\hat{\tau} \doteq \argmax_{\tilde{\tau} \in \mathcal{T}}
\sum_{\tau \in \mathcal{T}} G\big(\omega(\tilde{\tau}), \omega(\tau)\big)
\pi_{\tau}.
\]
When comparing two indicator ranks, the gain function $G$ assigns zero gain to
each discrepancy between them, a unit gain to matched zeroes (true negatives)
and a gain of $\kappa > 0$ to matched ones (true positives). For example, if
$K = 7$, then $G(\omega(3), \omega(5)) = 2 + 3\kappa$ since there are three
matched ones from positions 1 through 3, two mismatches from positions 4 and
5, and two matched zeros from the last two positions, 6 and 7. Thus,
\[
G\big(\omega(\tau_1), \omega(\tau_2)\big) = K - \max\{\tau_1, \tau_2\} +
\kappa\min\{\tau_1, \tau_2\}.
\]

Then,
\[
\begin{split}
\hat{\tau} &= \argmax_{\tilde{\tau} \in \mathcal{T}} \sum_{\tau \in \mathcal{T}}
\Big( \kappa \min\{\tau, \tilde{\tau}\} - \max\{\tau, \tilde{\tau}\} \Big)
\pi_\tau \\
&= \argmax_{\tilde{\tau} \in \mathcal{T}} \Bigg\{
\sum_{\tau \leq \tilde{\tau}} (\kappa \tau - \tilde{\tau}) \pi_\tau
+ \sum_{\tau > \tilde{\tau}} (\kappa \tilde{\tau} - \tau) \pi_\tau \Bigg\} \\
&= \argmax_{\tilde{\tau} \in \mathcal{T}} \Bigg\{
\sum_{\tau \leq \tilde{\tau}} ((\kappa + 1) \tau - \tilde{\tau}) \pi_\tau
+ \sum_{\tau > \tilde{\tau}} \kappa \tilde{\tau} \pi_\tau \Bigg\} \\
&= \argmax_{\tilde{\tau} \in \mathcal{T}} \Bigg\{
(\kappa + 1) \sum_{\tau \leq \tilde{\tau}} \tau \pi_\tau
- \tilde{\tau} \Pr(\tau \leq \tilde{\tau} \given Y)
+ \kappa \tilde{\tau} \big(1 - \Pr(\tau \leq \tilde{\tau} \given Y)\big) \Bigg\} \\
&= \argmax_{\tilde{\tau} \in \mathcal{T}} \Big\{
(\kappa + 1) \Exp\big[\tau \given \tau \leq \tilde{\tau}, Y\big]
+ \tilde{\tau} \big[\kappa - (\kappa + 1)
  \Pr(\tau \leq \tilde{\tau} \given Y)\big] \Big\},
\end{split}
\]
that is, $\hat{\tau} = \argmax_{\tilde{\tau} \in \mathcal{T}}
g(\tilde{\tau})$, where $g$ is last expression within brackets above. Clearly,
$g(0) = 0$; in general,
\[
g(j) = (\kappa + 1)\sum_{i=0}^j i \pi_i +
j \big[\kappa - (\kappa + 1) \sum_{i=0}^j \pi_i\big] \\
= \kappa j +
(\kappa + 1) \underbrace{\sum_{i=0}^j (j - i) \pi_i}_{\doteq s_j}.
\]
But since $s_j = j \sum_{i=0}^j \pi_i - \sum_{i=0}^j i \pi_i$, it follows that
\[
s_{j+1} = (j + 1)(\sum_{i=0}^j \pi_i + \pi_{j+1}) - \sum_{i=0}^j i \pi_i -
(j+1)\pi_{j+1} = s_j + \sum_{i=0}^j \pi_i.
\]
Thus,
\[
g(j + 1) = \kappa (j + 1) - (\kappa + 1)\Big(s_j + \sum_{i=0}^j \pi_i\Big)
= g(j) + \kappa - (\kappa + 1) \sum_{i=0}^j \pi_i,
\]
and so $g(j+1) > g(j)$ if and only if $\kappa - (\kappa+1)\sum_{i=0}^j \pi_i$,
that is, $\kappa > \sum_{i=0}^j \pi_i / (1 - \sum_{i=0}^j \pi_i)$, when
$\kappa$ exceeds the cumulative odds. Thus, since $\sum_{i=0}^j \pi_i$ is
non-decreasing, we conclude that
\[
\hat{\tau} = \max\Bigg\{\tilde{\tau} \in \mathcal{T} \,:\,
\sum_{\tau=0}^{\tilde{\tau}} \pi_{\tau} < \frac{\kappa}{1 + \kappa}\Bigg\},
\]
so we propose to expand the basis expansion up to when the cumulative
posterior exceeds the $\kappa/(1+\kappa)$ threshold.

\subsection{ECM Specifics: Spike and Slab Variable Selection}
\label{ssec:ecm}
For a particular $\theta_j$ and $\tau_j$ we have:
\[
Y \given \theta, \tau \ind \text {\sf Po} \big(\exp(D_{X_j}(v)^\top\theta) \big)
\]
and $\theta \given \tau \sim N\Big(0, \Sigma \Big)$ where
$\Sigma = \lambda^{-1} M_\tau^{1/2} {(D_{X_j}^\top L_w D_{X_j})}^{-}M_\tau^{1/2}$
and $M_\tau = \text{Diag}_{i=1\; .\;.\;.\; K}
\{\text{I}(i > \tau)V_0 + \text{I}(i \leq \tau)\}.$ $\tau$ is as defined in
\eqref{eq:tau}. We wish to optimize the expected log joint~\citep{dempster1977}:
\begin{equation}
Q(\theta;\theta^{(t)}) = \Exp_{\tau \given Y;\theta^{(t)}}[\log\Pr(\theta,\tau \given Y)].
\label{eq:Q1}
\end{equation}
Thus, for the E-step we need,
\[
\nu_i^{(t)} = \Exp[\text{I}(i > \tau)] = \sum_{l=1}^{i-1}\Pr(\tau = l \given Y, \theta^{(t)})
\]
where
\[
\Pr(\tau = l \given Y, \theta^{(t)}) =
\frac{\Pr(\theta^{(t)} \given \tau = l) \Pr(\tau = l)}%
{\sum_{l = 1}^K \Pr(\theta^{(t)} \given \tau = l) \Pr(\tau = l)}.
\]
Next we update $\theta$ by maximizing the expected log likelihood given in~\eqref{eq:Q1}.
\begin{multline*}
Q(\theta;\theta^{(t)})  = c - v\text{exp}(D_{X_j}(v)^\top\theta^{(t)})+\sum_{v}Y_v (D_{X_j}(v)^\top\theta^{(t)}) - \\  \frac{\lambda}{2} \Exp[{\theta^{(t)}}^\top M_\tau^{-1/2} (D_{X_j}^\top L_w D_{X_j}) M_\tau^{-1/2} \theta^{(t)}].
\end{multline*}
We see that updating $\theta$ is equivalent to fitting a Poisson regression
with prior precision on $\theta$.  More specifically, we have
\begin{equation*}
\Exp[{\theta^{(t)}}^\top M_\tau^{-1/2} (D_{X_j}^\top L_w D_{X_j}) M_\tau^{-1/2} \theta^{(t)}] =  {\theta^{(t)}}^\top [T(v^{(t)}) \circ (D_{X_j}^\top L_w D_{X_j}] \theta^{(t)}
\end{equation*}
where $\circ$ is the Hadamard (element-wise) product and
\begin{equation*}
T(v) = \big[1-\nu_{max\{i,k\}}+(\nu_{max\{i,k\}}-\nu_{min\{i,k\}})V_0^{-\frac{1}{2}}+\nu_{min\{i,k\}}V_0^{-1}\big]_{i,k=1,\dots,K}.
\end{equation*}

\section{Model Inference}
\subsection{EM Specifics: Finding the MAP Coefficient Estimates} \label{hotzone}
Let $\delta$ be a vector of the current parameters: $\zeta$, $\omega$, and
$\theta$.  We have: $\Pr (Y \given \delta) = \sum_z \Pr (Y,Z \given \delta)$,
with $Y_v \given Z_v, \delta \ind \text {\sf Po}
\big(\exp(Z_v\zeta+(1-Z_v)D_X(v)^\top\theta) \big)$ and
$Z_v|\delta \ind \text{\sf Bern} \big( \text{logit}^{-1}(D_U(v)^\top\omega)\big)$.
Again, we wish to maximize the expected log joint:
\begin{equation}
\begin{split}
Q(\delta;\delta^{(t)}) &=
\Exp_{Z \given Y;\delta^{(t)}}[\log\Pr(Y,Z \given \delta)]\\
&= \underbrace{\Exp_{Z \given Y;\delta^{(t)}}[\log\Pr(Y \given Z, \delta)]}_{Q_1}
+ \underbrace{\Exp_{Z \given Y;\delta^{(t)}}[\log\Pr(Z \given \delta)}_{Q_2}].
\end{split}
\label{eq:exp}
\end{equation}
For the E-step we need
$\pi_v^{(t)} \doteq \Exp_{Z \given Y;\delta^{(t)}} [Z_v]$, that is,
\[
\pi_v^{(t)} = \frac{\Pr(Y_v \given Z_v = 1, \theta^{(t)})
      \Pr(Z_v = 1 \given \omega^{(t)})}
    {\sum_{\tilde{Z}_v \in \{0,1\}}
          \Pr(Y_v \given \tilde{Z}_v, \theta^{(t)}, \zeta^{(t)})
          \Pr(\tilde{Z}_v \given \omega^{(t)})}.
\]
It follows that $\pi_v^{(t)} = \gamma_v(\omega^{(t)}, \zeta^{(t)},
\theta^{(t)})$ with $\gamma_v$ as in~\ref{eq:gamma}.  Next, we update $\zeta$,
$\theta$, and $\omega$ by maximizing the expected log likelihood given
in~\eqref{eq:exp}.  From the first part:
\begin{align*}
Q_1 = & \Exp_{Z|Y;\delta^{(t)}} \Big[ \sum_v -\exp \big( -Z_v\zeta^{(t)} +
(1-Z_v)D_X(v)^\top\theta^{(t)} \big) + 
Y_v \big( Z_v\zeta^{(t)}+(1-Z_v)D_X(v)^\top\theta^{(t)} \big) \Big]\\
= & \Exp_{Z \given Y; \delta^{(t)}} \Big[ \sum_v Z_v \big( Y_v\zeta^{(t)} -
\exp(\zeta^{(t)}) \big) + (1-Z_v) \big( Y_v D_X(v)^\top \theta^{(t)} -\exp(D_X(v)^\top\theta^{(t)}) \big) \Big]\\
= &\sum_v \pi^{(t)}_v\big(Y_v\zeta^{(t)}-\exp(\zeta^{(t)})\big) +
(1-\pi_v^{(t)})\big(Y_v D_X (v)^\top \theta^{(t)}
-\exp(D_X(v)^\top\theta^{(t)})\big) \\
= & \sum_v\pi_v^{(t)}Y_v (\zeta^{(t)}+\log\pi_v^{(t)}) -
\pi_v^{(t)}Y_v\log\pi_v^{(t)} - \exp(\zeta^{(t)}+\log\pi_v^{(t)})+ 
(1-\pi_v^{(t)})Y_v \big(D_X(v)^\top\theta^{(t)} \\
&\qquad + \log(1-\pi_v^{(t)})\big) -
(1-\pi_v^{(t)})Y_v\log(1-\pi_v^{(t)}) - 
\exp\big(D_X(v)^\top\theta^{(t)}+\log(1-\pi_v^{(t)})\big).
\end{align*}
Analyzing the terms that contain $\zeta$, we see that updating $\zeta$ is
equivalent to fitting a quasi-Poisson regression with non-integer response,
$\pi^{(t)}Y$.  We have $\pi^{(t)}Y \sim \text{Quasi-Po} \Big[
\exp(\zeta+\log\pi^{(t)}) \Big]$. Similarly, we update $\theta$
where $(1-\pi^{(t)})Y \sim \text{Quasi-Po} \Big[ \exp\big(D_X
\theta+\log(1-\pi^{(t)})\big)\Big]$ and we use prior precision
$\lambda_{\theta} D_X^\top L_{w(\psi)}D_x.$  Now, from the second part:
\begin{align*}
Q_2 = & \Exp_{Z \given Y;\delta^{(t)}} \Big[ Z_v \log\big(\text{logit}^{-1}(D_U(v)^\top\omega^{(t)}) \big)+ 
(1-Z_v) \Big( 1-\log\big( \text{logit}^{-1}(D_U(v)^\top\omega^{(t)})\big) \Big) \Big ]\\
= & \sum_v\pi_v^{(t)}\log\big( \text{logit}^{-1}(D_U(v)^\top\omega^{(t)}) \big)+ 
(1-\pi_v^{(t)}) \Big( 1-\log\big( \text{logit}^{-1}(D_U(v)^\top\omega^{(t)})\big) \Big).
\end{align*}
Using similar reasoning as in the previous step, we update $\omega$ using a
quasi-Bernoulli regression.  That is, $\pi^{(t)} \sim \text{Quasi-Bern} \big[
\text{logit}^{-1} (D_U\omega ) \big].$ We use prior precision $\lambda_\omega
D_U^\top L_{w(\psi)} D_U$.

\newpage

\subsection{Sampling Algorithm} \label{sampling}
Given initial estimates found from the EM algorithm presented in the preceding
section, we now turn our attention to the Gibbs sampler and the conditional
distributions in~\eqref{eq:conditional}.
Letting $\Sigma = \lambda_{\theta}^{-1} (D_X^\top L_{w(\psi)}D_x)^-$ we have
for $\zeta$ and $\theta$
\begin{equation*}
 \Pr(\zeta, \theta \given \omega, Y_v, Z_v) \propto \prod_v
 \text{\sf Po}\Big[\exp \big(Z_v \zeta + (1-Z_v)({D_X(v)}^\top \theta)\big)\Big]
 \text{\sf N}(0, \Sigma).
\end{equation*}
It follows, similarly to Section~\ref{hotzone}, that up to a constant,
\begin{align*}
\log \Pr(\zeta, \theta \given \omega, Y_v, Z_v) = & 
\sum_v -\exp \big(Z_v\zeta + (1-Z_v)D_X(v)^\top\theta \big) + Y_v \big( Z_v\zeta + (1-Z_v)D_X(v)^\top\theta \big) -  \frac{1}{2}\theta^\top \Sigma^{-1}\theta \\
= & \sum_v Z_v \big( Y_v\zeta - \exp(\zeta) \big) +
(1-Z_v) \big( Y_v D_X(v)^\top \theta -
\exp(D_X(v)^\top\theta) \big) - \frac{1}{2}\theta^\top\Sigma^{-1}\theta\\
%= & \sum_{v, z = 1} \underbrace{-\exp(\zeta) + Y_v\zeta}_{1} + \sum_{v, z = 0} \underbrace{-\exp(D_X(v)^\top\theta) + Y_v D_X(v)^\top\theta -}_{2} \\
%& \underbrace{\frac{1}{2}\theta^\top\Sigma^{-1}\theta.}_{2}
\end{align*}
and so, isolating the terms for $Z_v = 0$ and $Z_v = 1$ containing $\theta$ and
$\zeta$ respectively, we see that
\begin{equation*}
Y_v \given Z_v = 0 \ind \text{\sf Po}(\exp {D_X(v)}^\top \theta )
\qquad \text{and} \qquad
Y_v \given Z_v = 1 \ind \text{\sf Po}(e^\zeta),
\end{equation*}
with prior distribution on $\theta$ described in~\eqref{eq:model}.
An analogous derivation shows that
\[
Z_v \ind \text{\sf Bern}(\text{logit}^{-1}D_U(v)^\top\omega).
\]
That is, sampling $\zeta, \theta$ and $\omega$ is equivalent to sampling from
the posterior distribution of a GLM with a multivariate normal prior on the
coefficients $\theta$ and $\omega$. Sampling is done using the Riemannian
manifold Metropolis adjusted Langevin algorithm (MALA), as described
in~\citet{sampling2}.  The proposal mechanism takes into account the natural
geometry of the target density and makes transition proposals that are informed
by its local structure.  Specifically, the proposal density is a normal
distribution, with mean based on the curvature of the space along the direction
of steepest gradient and covariance given by the scaled inverse Fisher
information matrix of the posterior.  The process, in the general case, is as
follows:
\begin{enumerate}
\item Set $t = 1$ and begin with $\beta = \beta^{(0)};$
\item Sample $\beta^*$ from proposal density, $q$;
\item Accept with probability, $\rho(\beta^{(t-1)},\beta^*).$  If accept, $\beta^{(t)} = \beta^*$; if reject, $\beta^{(t)} = \beta^ {(t-1)};$
\item Set t = t + 1 and return to step 2;
\end{enumerate}
where
\begin{equation*}
\rho(\beta^{(t-1)},\beta^*) = \textrm{min}\bigg(1, \frac{\pi(\beta^*)q(\beta^*,\beta^{(t-1)})}{\pi(\beta^{(t-1)}q(\beta^{(t-1)},\beta^*)}\bigg).
\end{equation*}
Here $q(\beta^*|\beta^n,\epsilon) = N\big(\beta^*|\,\mu(\beta^n,\epsilon), \epsilon^2G^{-1}(\beta^n)\big)$ where
 \begin{align*}
 \mu(\beta^n,\epsilon)_i &= \beta^n_i + \frac{\epsilon^2}{2}\{G^{-1}(\beta^n)\nabla_\beta\textrm{log}\{p(\beta^n)\}\}_i - \epsilon^2\sum_{j=1}^D \big \{G^{-1}(\beta^n) \frac{\partial G(\beta^n)}{\partial \beta_j}G^{-1}(\beta^n)\big \}_{ij} + \\
 & \frac{\epsilon^2}{2}\sum_{j=1}^D \big \{G^{-1}(\beta^n)\big \}_{ij} \textrm{tr}\Bigg \{ G^{-1}(\beta^n)\frac{\partial G(\beta^n)}{\partial \beta_j}\Bigg \},
 \end{align*}
$\epsilon$ is a step parameter, and $G$ is the Fisher information matrix of the posterior distribution.       That is, with $\Sigma$ defined as above, $G(\beta) = \Exp[-\frac{\partial^2 \Pr(\beta|y) }{\partial \beta \partial \beta^\top}] = X^\top \Omega(\beta)X + \Sigma^{-1}$ and $\frac{\partial G}{\partial \beta_{i}} = X^\top\Omega(\beta)V^iX.$  In the logistic model, given the logit link function:
\begin{itemize}
\item $\Omega(\omega)_{v,v} = \text{logit}^{-1}(D_U(v)^\top \omega)\big(1-\text{logit}^{-1}(D_U(v)^\top \omega)\big)$
\item $V^i = \big(1-2\text{logit}^{-1}(D_U(v)^\top \omega)\big)D_U(v)_i$
\end{itemize}
in accordance with the example in Section 7 of~\citep{sampling2}.  For the Poisson model, with the log link function, we have:
\begin{itemize}
\item $\Omega(\theta)_{v,v} = \text{exp}(D_X(v)^\top\theta)$
\item $V^i = D_X(v)_i.$
\end{itemize}
We use a step size based on the recommendation of~\citet{sampling3} on the order of $D^{-1/3}$ where $D$ is the dimension of the parameter space.  In practice, this resulted in an acceptance rate of approximately $75\%$.

\newpage

\subsection{Posterior distribution of linear predictors based on Gibbs Sampling}
\label{samplingresults}
\begin{figure}[h!]
\centering
\includegraphics[scale=.5]{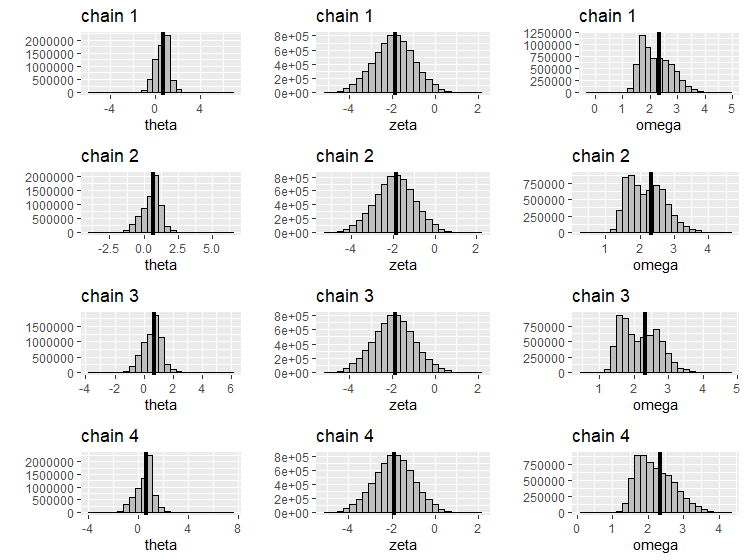}
\caption{The linear predictors (i.e. $D_X\theta, X\zeta, and D_U\omega$).  Per the sampling scheme, there are 1000 values for each residential intersection in Boston. }
\label{fig:Gibbs}
\end{figure}

Here we plot the posterior distributions of the linear predictors and
superimpose the results from the EM algorithm point estimates; they are in
agreement.  The analysis was run on a MacBook Pro laptop, with an Apple M2 chip
featuring 8 cores, comprising 4 performance and 4 efficiency cores. The system
is outfitted with 16 GB of memory and no parallelization scheme was
implemented.  As a point of comparison, the EM algorithm ran quite quickly
($\sim 21$ minutes), but running the entire Gibbs sampling scheme as described
in the paper (4 chains) took $\sim 4$ days.  We leave the choice to the
practitioner.  The network is composed of 12,763 nodes (i.e. street segments).

\newpage
\section{Boston Neighborhoods}
\begin{figure}[h!]
\centering
\includegraphics[scale=.4]{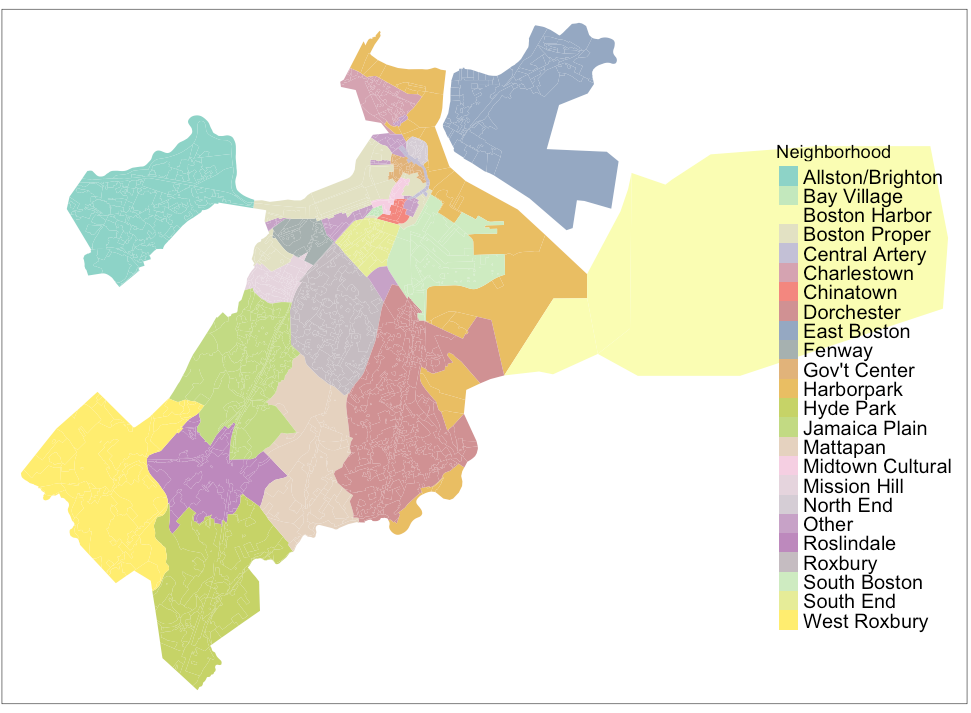}
\caption{Neighborhood designations in Boston, MA. }
\label{fig:Neighborhoods}
\end{figure}

\label{lastpage}

%\bibliographystyle{imsart-nameyear} % Style BST file
%\bibliography{burglary2}       % Bibliography file (usually '*.bib')

\bibliography{burglary2}

@book{banerjee2014,
 author               = {Banerjee, Sudipto and Carlin, Bradley P and Gelfand, Alan E},
 publisher            = {{CRC} Press},
 title                = {Hierarchical Modeling and Analysis for Spatial Data},
 year                 = {2014},
 }

@inproceedings{belkin2004,
 author               = {Belkin, Mikhail and Matveeva, Irina and Niyogi, Partha},
 booktitle            = {International Conference on Computational Learning Theory},
 organization         = {Springer},
 pages                = {624--638},
 title                = {Regularization and Semi-Supervised Learning on Large Graphs},
 year                 = {2004},
 }

@article{bernasco2009,
 author               = {Bernasco, Wim and Block, Richard},
 journal              = {Criminology},
 number               = {1},
 pages                = {93--130},
 publisher            = {Wiley Online Library},
 title                = {Where Offenders Choose to Attack: A Discrete Choice Model of Robberies in {C}hicago},
 volume               = {47},
 year                 = {2009},
 }

@book{cormen2001,
 author               = {Cormen, Thomas H and Leiserson, Charles Eric and Rivest, Ronald L and Stein, Clifford},
 publisher            = {MIT press Cambridge},
 title                = {Introduction to Algorithms},
 volume               = {6},
 year                 = {2001},
 }

@article{dempster1977,
 author               = {Dempster, Arthur P and Laird, Nan M and Rubin, Donald B},
 journal              = {Journal of the Royal Statistical Society Series B},
 number               = {1},
 pages                = {1--38},
 publisher            = {Wiley},
 title                = {Maximum Likelihood from Incomplete Data via the {EM} Algorithm},
 volume               = {39},
 year                 = {1977},
 }

@techreport{eck2005,
 author               = {Eck, John and Chainey, Spencer and Cameron, James and Wilson, R},
 institution          = {National Institute of Justice},
 title                = {Mapping Crime: Understanding Hotspots},
 year                 = {2005},
 }

@book{garner2001,
 author               = {Garner, Bryan A},
 publisher            = {Oxford University Press, USA},
 title                = {A Dictionary of Modern Legal Usage},
 year                 = {2001},
 }

@book{kolaczyk2009,
 author               = {Kolaczyk, Eric D},
 publisher            = {Springer},
 title                = {Statistical Analysis of Network Data},
 year                 = {2009},
 }

@book{kolaczyk2014,
 author               = {Kolaczyk, Eric D and Cs{\'a}rdi, G{\'a}bor},
 publisher            = {Springer},
 title                = {Statistical Analysis of Network Data with {R}},
 year                 = {2014},
 }

@article{lanckriet2004,
 author               = {Lanckriet, Gert RG and De Bie, Tijl and Cristianini, Nello and Jordan, Michael I and Noble, William Stafford},
 journal              = {Bioinformatics},
 number               = {16},
 pages                = {2626--2635},
 publisher            = {Oxford Univ Press},
 title                = {A Statistical Framework for Genomic Data Fusion},
 volume               = {20},
 year                 = {2004},
 }

@article{li2008,
 author               = {Li, Caiyan and Li, Hongzhe},
 journal              = {Bioinformatics},
 number               = {9},
 pages                = {1175--1182},
 publisher            = {Oxford Univ Press},
 title                = {Network-Constrained Regularization and Variable Selection for Analysis of Genomic Data},
 volume               = {24},
 year                 = {2008},
 }

@book{mccullagh1989,
 author               = {McCullagh, P and Nelder, J A},
 publisher            = {Chapman \& Hall},
 title                = {Generalized Linear Models},
 year                 = {1989},
 }

@book{ramsay2005,
 author               = {Ramsay, J and Silverman, B W},
 publisher            = {Springer},
 title                = {Functional Data Analysis},
 year                 = {2005},
 }

@incollection{smola2003,
 author               = {Smola, Alexander J and Kondor, Risi},
 booktitle            = {Learning Theory and Kernel Machines},
 pages                = {144--158},
 publisher            = {Springer},
 title                = {Kernels and Regularization on Graphs},
 year                 = {2003},
 }

@misc{opendata,
author 				= {{Open Data}},
  title 			= {{Boston Maps}},
  howpublished 		= {\url{http://bostonopendata-boston.opendata.arcgis.com/}},
  note 				= {Retrieved February 16, 2016},
  year				= {2016}
}

@misc{bostondata,
  author		    = {{City of Boston}},
  title 			= {{Data Boston}},
  howpublished 		= {\url{https://data.cityofboston.gov/}},
  note 				= {Retrieved February 16, 2016},
  year				= {2016}
}

@misc{snapnets,
  author       = {Leskovec, Jure and Krevl, Andrej},
  title        = {{SNAP Datasets: Stanford Large Network Dataset   Collection}},
  howpublished = {\url{http://snap.stanford.edu/data}},
  year		   = {2014},
  note 		   = {Retrieved March 23, 2017}
}

@article{george1993,
  title={Variable selection via Gibbs sampling},
  author={George, Edward I and McCulloch, Robert E},
  journal={Journal of the American Statistical Association},
  volume={88},
  number={423},
  pages={881--889},
  year={1993},
  publisher={Taylor \& Francis Group}
}

@article{emvs,
  title={EMVS: The EM approach to Bayesian variable selection},
  author={Ro{\v{c}}kov{\'a}, Veronika and George, Edward I},
  journal={Journal of the American Statistical Association},
  volume={109},
  number={506},
  pages={828--846},
  year={2014},
  publisher={Taylor \& Francis}
}

@article{levina2016,
  title={Prediction models for network-linked data},
  author={Li, Tianxi and Levina, Elizaveta and Zhu, Ji},
  journal={The Annals of Applied Statistics},
  volume={13},
  number={1},
  pages={132--164},
  year={2019},
  publisher={Institute of Mathematical Statistics}
}

@book{monte,
  title={Monte Carlo statistical methods},
  author={Robert, Christian and Casella, George},
  year={2013},
  publisher={Springer Science \& Business Media}
}

@article{sampling,
  title={Sampling from the posterior distribution in generalized linear mixed models},
  author={Gamerman, Dani},
  journal={Statistics and Computing},
  volume={7},
  number={1},
  pages={57--68},
  year={1997},
  publisher={Springer}
}

@article{sampling2,
  title={Riemann manifold langevin and hamiltonian monte carlo methods},
  author={Girolami, Mark and Calderhead, Ben},
  journal={Journal of the Royal Statistical Society: Series B (Statistical Methodology)},
  volume={73},
  number={2},
  pages={123--214},
  year={2011},
  publisher={Wiley Online Library}
}

@article{sampling3,
  title={Optimal scaling of discrete approximations to Langevin diffusions},
  author={Roberts, Gareth O and Rosenthal, Jeffrey S},
  journal={Journal of the Royal Statistical Society: Series B (Statistical Methodology)},
  volume={60},
  number={1},
  pages={255--268},
  year={1998},
  publisher={Wiley Online Library}
}

@article{mengrubin93,
  title={Maximum likelihood estimation via the {ECM} algorithm: A general framework},
  author={Meng, Xiao-Li and Rubin, Donald B},
  journal={Biometrika},
  volume={80},
  number={2},
  pages={267--278},
  year={1993},
  publisher={Biometrika Trust}
}

@article{bowers,
  title={Prospective hot-spotting: the future of crime mapping?},
  author={Bowers, Kate J and Johnson, Shane D and Pease, Ken},
  journal={British journal of criminology},
  volume={44},
  number={5},
  pages={641--658},
  year={2004},
  publisher={Oxford University Press}
}

@article{mohler,
  title={Self-exciting point process modeling of crime},
  author={Mohler, George O and Short, Martin B and Brantingham, P Jeffrey and Schoenberg, Frederic Paik and Tita, George E},
  journal={Journal of the American Statistical Association},
  volume={106},
  number={493},
  pages={100--108},
  year={2011},
  publisher={Taylor \& Francis}
}

@article{balocchi,
  title={Spatial modeling of trends in crime over time in Philadelphia},
  author={Balocchi, Cecilia and Jensen, Shane T},
  journal={The Annals of Applied Statistics},
  volume={13},
  number={4},
  pages={2235--2259},
  year={2019},
  publisher={Institute of Mathematical Statistics}
}

@article{rhat,
  title={Rank-normalization, folding, and localization: An improved R-hat for assessing convergence of MCMC},
  author={Vehtari, Aki and Gelman, Andrew and Simpson, Daniel and Carpenter, Bob and B{\"u}rkner, Paul-Christian},
  journal={arXiv preprint arXiv:1903.08008},
  year={2019}
}

@article{carbon,
  title={Carbon Free Boston: Social equity report 2019},
  author={Cleveland, Cutler and Stanton, Liz and Woods, Bryndis and Martin, Atyia and Fortune, D’Janapha and Walsh, Michael and Castigliego, Joshua and Perez, Taylor and Galante, Emma and others},
  year={2019},
  publisher={Boston University Institute for Sustainable Energy}
}

@techreport{gentrification,
  title={Gentrification and the amenity value of crime reductions: Evidence from rent deregulation},
  author={Palmer, Christopher J and Pathak, Parag A and others},
  year={2017},
  institution={National Bureau of Economic Research}
}

@article{kron,
  title={Kron reduction of graphs with applications to electrical networks},
  author={Dorfler, Florian and Bullo, Francesco},
  journal={IEEE Transactions on Circuits and Systems I: Regular Papers},
  volume={60},
  number={1},
  pages={150--163},
  year={2012},
  publisher={IEEE}
}

@article{davies,
  title={Examining the relationship between road structure and burglary risk via quantitative network analysis},
  author={Davies, Toby and Johnson, Shane D},
  journal={Journal of Quantitative Criminology},
  volume={31},
  pages={481--507},
  year={2015},
  publisher={Springer}
}

@article{dual,
  title={The network analysis of urban streets: A dual approach},
  author={Porta, Sergio and Crucitti, Paolo and Latora, Vito},
  journal={Physica A: Statistical Mechanics and its Applications},
  volume={369},
  number={2},
  pages={853--866},
  year={2006},
  publisher={Elsevier}
}

@article{forecasting,
  title={Network analysis of city streets: forecasting burglary risk in small areas},
  author={Mahfoud, Maria and Bhulai, Sandjai and van der Mei, Rob and Erkin, Dimitry and Dugundji, Elenna},
  journal={International Journal On Advances in Security},
  year={2019}
}

@article{frith,
  title={ROLE OF THE STREET NETWORK IN BURGLARS'SPATIAL DECISION-MAKING},
  author={Frith, Michael J and Johnson, Shane D and Fry, Hannah M},
  journal={Criminology},
  volume={55},
  number={2},
  pages={344--376},
  year={2017},
  publisher={Wiley Online Library}
}

@misc{displacement,
  author		    = {{City of Boston}},
  title 			= {{Boston Housing Conditions and Real Estate Trends Report}},
  howpublished 		= {\url{https://www.bostonplans.org/getattachment/066b23c5-cab9-4731-a338-f6e57e3ef55f}},
  note 				= {Retrieved January 8, 2023},
  year				= {2022}
}

@inproceedings{kim2018,
  title={Crime analysis through machine learning},
  author={Kim, Suhong and Joshi, Param and Kalsi, Parminder Singh and Taheri, Pooya},
  booktitle={2018 IEEE 9th Annual Information Technology, Electronics and Mobile Communication Conference (IEMCON)},
  pages={415--420},
  year={2018},
  organization={IEEE}
}

@article{predictive,
  title={Predictive policing: Review of benefits and drawbacks},
  author={Meijer, Albert and Wessels, Martijn},
  journal={International Journal of Public Administration},
  volume={42},
  number={12},
  pages={1031--1039},
  year={2019},
  publisher={Taylor \& Francis}
}

@article{prevent,
  title={Crime displacement: what we know, what we don’t know, and what it means for crime reduction},
  author={Johnson, Shane D and Guerette, Rob T and Bowers, Kate},
  journal={Journal of Experimental Criminology},
  volume={10},
  pages={549--571},
  year={2014},
  publisher={Springer}
}

@article{besag,
  title={Bayesian image restoration, with two applications in spatial statistics},
  author={Besag, Julian and York, Jeremy and Molli{\'e}, Annie},
  journal={Annals of the institute of statistical mathematics},
  volume={43},
  pages={1--20},
  year={1991},
  publisher={Springer}
}
\bibliographystyle{agsm}

\end{document}